\documentclass[aps,prl,floats,twocolumn,showpacs,superscriptaddress]{revtex4-1}
\usepackage{graphicx}
\usepackage{amssymb}
\usepackage{color}
\begin{document}
%\title{Emergence of neuronal spike correlations by the interplay of weak noise and modulation}
\title{Emergence of Spike Correlations in Periodically Forced Excitable Systems}

\author{Jos\'{e} A. Reinoso}
\email{aparicioreinoso@gmail.com} 
\affiliation{Departament de Fisica, Universitat Politecnica de Catalunya, Colom 11, ES-08222 Terrassa, Barcelona,
Spain}
\author{M.C. Torrent}
\email{carme.torrent@upc.edu}
\affiliation{Departament de Fisica, Universitat Politecnica de Catalunya, Colom 11, ES-08222 Terrassa, Barcelona, Spain}
\author{Cristina Masoller}
\email{cristina.masoller@upc.edu} 
\affiliation{Departament de Fisica, Universitat Politecnica de Catalunya, Colom
11, ES-08222 Terrassa, Barcelona, Spain} 

% revista: plos computational biology (impact 4.87)

\begin{abstract}
In sensory neurons the presence of noise can facilitate the detection of weak information-carrying signals, which are encoded and transmitted via correlated sequences of spikes. Here we investigate relative temporal order in spike sequences induced by a subthreshold periodic input, in the presence of white Gaussian noise. To simulate the spikes, we use the FitzHugh-Nagumo model, and to investigate the output sequence of inter-spike intervals (ISIs), we use the symbolic method of ordinal analysis. We find different types of relative temporal order, in the form of preferred ordinal patterns which depend on both, the strength of the noise and the period of the input signal. We also demonstrate a resonance-like behavior, as certain periods and noise levels enhance temporal ordering in the ISI sequence, maximizing the probability of the preferred patterns. Our findings could be relevant for understanding the mechanisms underlying temporal coding, by which single sensory neurons represent in spike sequences the information about weak periodic stimuli.
 %Time series analysis;  Models of single neurons and networks, Entropy and other measures of information

% 87.19.lo 	Information theory
%87.19.ln 	Oscillations and resonance
%87.19.lt 	Sensory systems

%keywords: neural coding, spike patterns, isi correlations, excitability, ordinal analysis, information theory, sensory systems, stochastic resonance
\end{abstract}
\date{\today}
\maketitle

\section{Introduction}

Many excitable systems, such as neurons and cardiac cells, display spiking output signals that can be analyzed by using an event-level approach, i.e., by detecting the times when the spikes occur, and then analyzing the statistics of the time intervals between successive spikes (inter-spike intervals, ISIs). Some important properties of ISI sequences are related to coherence and stochastic resonance phenomena. Coherence resonance refers to enhanced spike regularity under an optimal level of noise~\cite{Pikovsky:1997prl}, while stochastic resonance refers to enhanced detection and transmission of subthreshold time-varying signals, also under an optimal level of noise~\cite{Longtin:1993,collins_pre_1996,review0,review}. 

Another relevant property of ISI sequences is the presence of correlations \cite{longtin_1993, longtin_1997,nawrot_2007,lindner_jcn_2015}, which are known to influence the neuron's capacity of information transfer \cite{chacron_jns_2001,chacron_prl_2004, entropy_info_prl_1998, pre_2009}. In particular, while Gaussian white stochastic stimuli produce uncorrelated ISI sequences, correlated stochastic stimuli and information-carrying stimuli generate correlated spikes \cite{dante_1998,neiman_2001,middleton_pre_2003,neiman_pre}.

In the literature, temporal correlations in ISI sequences have been quantified by means of the serial correlation coefficients, $C_j$, 
\begin{equation}
 C_{j}=\frac{\left<\left(I_{i}-\left<I\right>\right) \left(I_{i-j}-\left<I\right>\right)\right>}{\sigma^{2}},
 \label{eqn:C}\end{equation}
where $j$ is an integer number, $\{\dots I_{i-1}, I_{i}, I_{i+1} \dots\}$ is the ISI sequence, and $\left<I\right>$ and $\sigma$ are the mean value and the standard deviation of the ISI distribution. 

Additional information can be gained by analyzing \textit{relative temporal ordering} in the ISI sequence, regardless of the values themselves. Relative temporal ordering can be quantified by computing the probabilities of \textit{ordinal patterns} (OPs) of length $L$, which are defined in terms of the order relations of $L$ data values \cite{bp_prl_2002}. For example, considering three consecutive ISIs, there are $L!=6$ possible order relations (neglecting equality) which are indicated in the inset of Fig. 2: $I_i<I_{i+1}<I_{i+2}$ gives pattern `012'; $I_{i+1}<I_{i}<I_{i+2}$ gives pattern `102', etc. If the six patterns are equally probable, one can conclude that there are no preferred order relations among three consecutive ISI values; in contrast, a non uniform distribution of OP probabilities reveals the presence of preferred and/or infrequent order relations. Longer order relations can be analyzed by either using lags (considering non-consecutive values, $I_{i}$, $I_{i+\tau}$ and $I_{i+2\tau}$) or by using longer patterns (for $L=4$ there are $4!=24$ possible order relations, for $L=5$ there are $5!=120$ order relations, etc.)

In general, the probabilities of the OPs and the serial correlation coefficients $C_j$ are related, but the OP probabilities are nontrivial functions of the $C_j$ coefficients, and \textit{vice versa}. In this sense, the information gained by using ordinal analysis complements that provided by correlation analysis. The OP probabilities are robust against observational noise and nonlinear perturbations \cite{bandt_2005}, and can be useful, for example, for contrasting the outputs of neuron models with observed ISI data; for fitting model parameters; for classifying different types of spike trains, etc.

Ordinal analysis has been extensively used to investigate biomedical signals and many other output signals of complex systems. It allows to classify different types of behaviors \cite{parlitz,special_issue}, to detect dynamical changes \cite{cao_pre_2004,njp_2015,prl_2016}, etc. (see \cite{review_zanin_entropy_2012,review_amigo} for many examples).  

Here our goal is to analyze order relations in ISI sequences generated by a single neuron driven by \textit{weak} periodic and stochastic inputs. We perform extensive simulations of the well-known FitzHugh-Nagumo (FHN) model driven by Gaussian white noise and a subthreshold sinusoidal input: without noise there are no spikes (but only subthreshold oscillations). The simulated ISI sequences are thus generated by the combined effects of noise and periodic forcing.

Without periodic forcing, the noise is uncorrelated and generates a sequence of statistically independent ISIs. In this situation, the six  $L=3$ OPs are equally probable. In contrast, when the neuron is also driven by a subthreshold periodic signal, temporal correlations emerge in the ISI sequence, which are detected and quantified by the probabilities of the six $L=3$ OPs. We demonstrate the presence of preferred OPs, which are tuned by i) the period of the input signal and ii) the strength of the noise. We also show that some OPs probabilities display the resonance-like feature of being enhanced for particular, signal-dependent noise levels. We conclude with a discussion of the relation between the OP probabilities, the mean ISI, and the serial correlation coefficients.

\section{Model}
\label{sec2}

The FHN equations are~\cite{Pikovsky:1997prl}:
\begin{eqnarray}
 \epsilon \frac{dx}{dt} &=& x-\frac{x^{3}}{3}-y, \\
\label{eqn:E2}
 \frac{dy}{dt}  &=& x+a+a_{o}\cos(2\pi t/T)+D\xi(t),
 \label{eqn:N2}
\end{eqnarray}
where $x$ is the fast variable and $y$ is the slow one, $\epsilon<<1$ and $a$ is a control parameter such that, when $|a|> 1$ there is a
stable node, and when $|a|< 1$, there is a stable limit cycle; $\xi(t)$ is a white Gaussian noise of zero mean  and
unit variance and $D$ is the noise strength; $a_o$ and $T$ are the amplitude and the period of the input signal.

\begin{figure}[h]
\includegraphics[width=6cm]{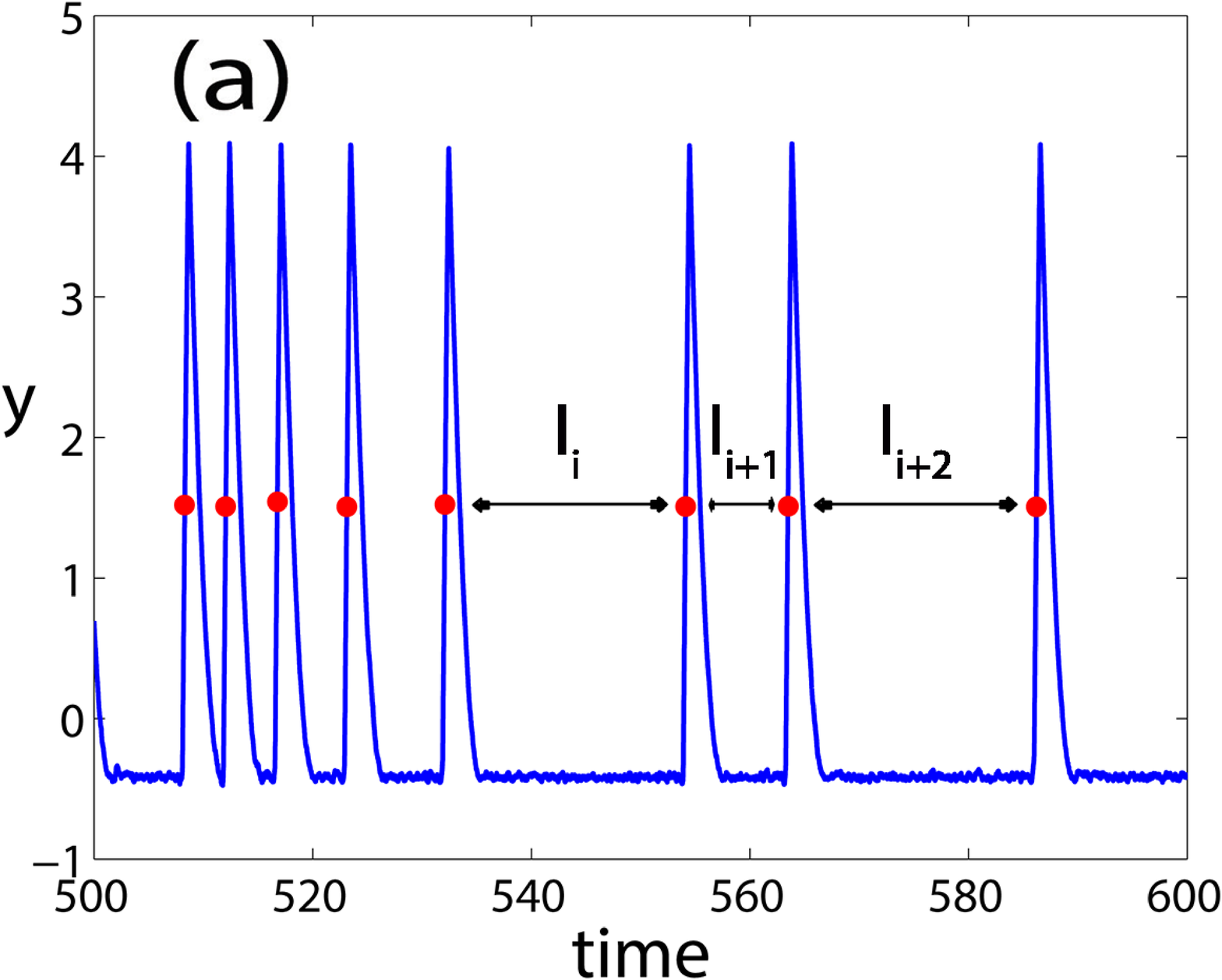}
\includegraphics[width=6cm]{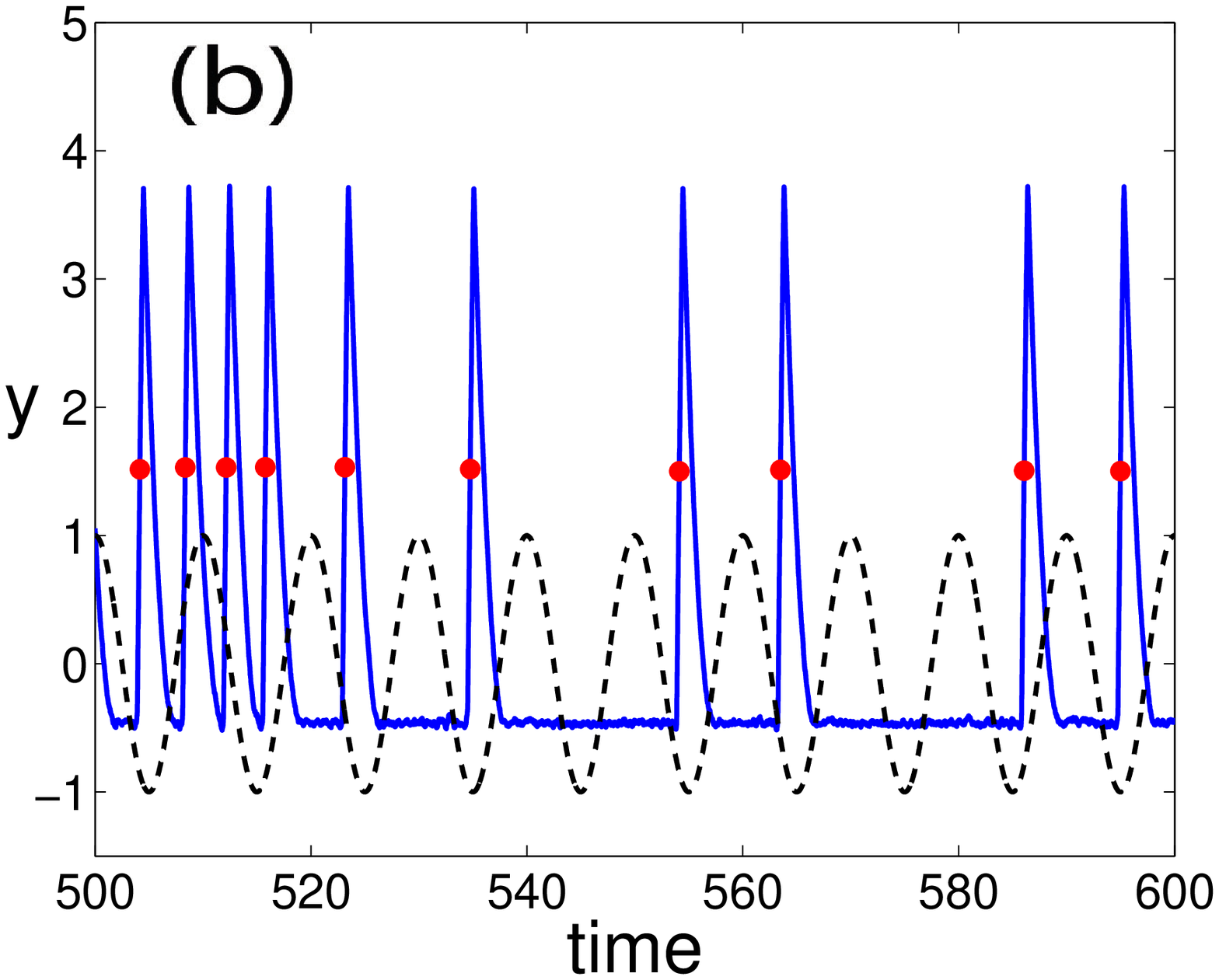}
\includegraphics[width=6cm]{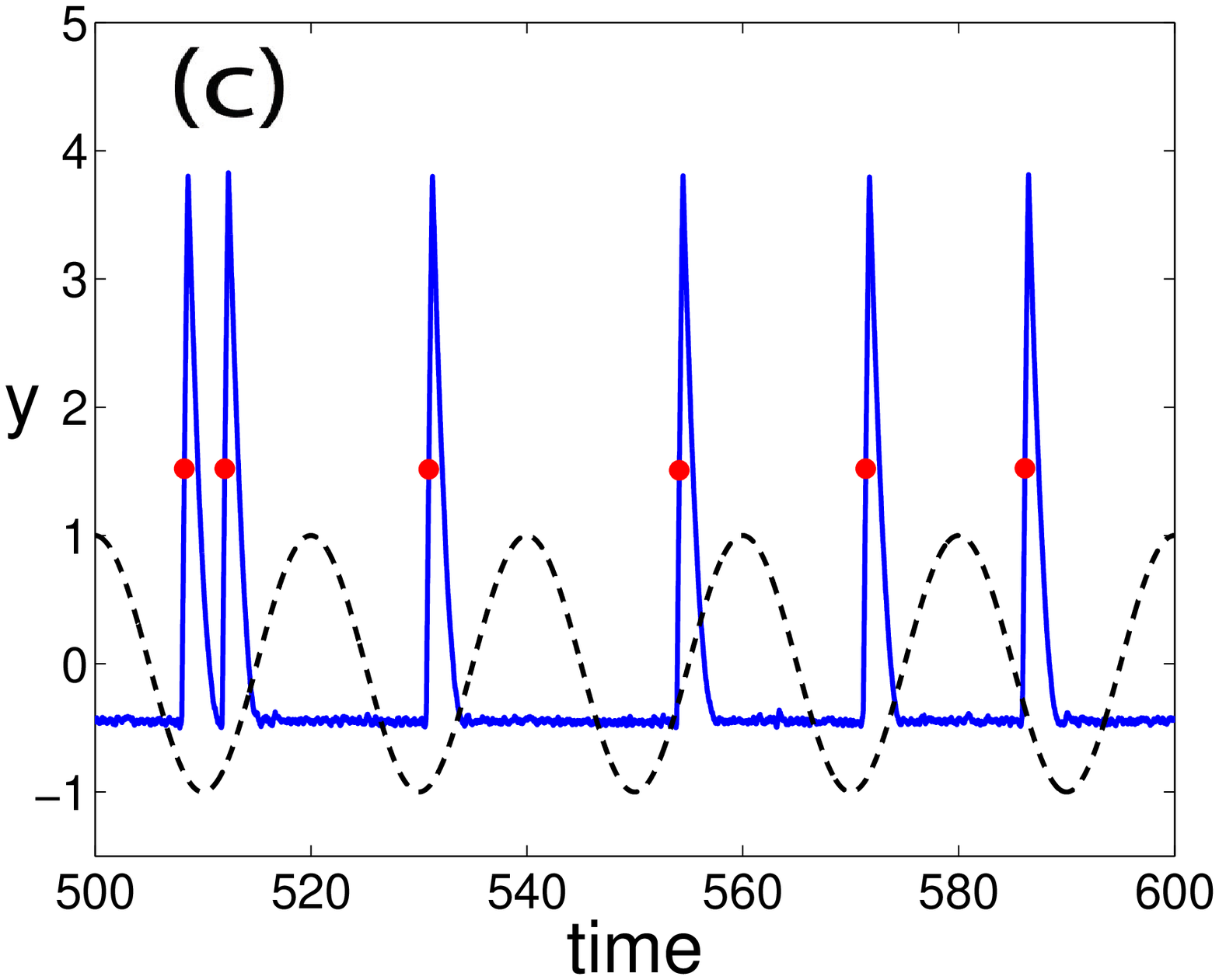}
\caption{(Color online) Time-series generated from the FHN model with parameters $a=1.05$, $\epsilon=0.01$, and $D=0.015$. In $(a)$ $a_{o}=0$, while in $(b)$ and $(c)$ $a_{o}=0.02$, $T=10$ and $T=20$ respectively. 
The spike times are detected with the threshold $y=1.5$. In panels $(b)$
and $(c)$ the dashed line indicates the value of $cos(2\pi t/T)$.
\label{fig:1}}
\end{figure}

The FHN model is simulated with parameters as in~\cite{Pikovsky:1997prl}: $a=1.05$ and $\epsilon=0.01$; $a_0$ and $T$ are varied such that the input signal is kept subthreshold (without noise there are no spikes). Figure~\ref{fig:1} displays typical spike sequences, where the spike times, $t_i$, are detected by using a threshold. Then, from the ISI sequence $\{I_i\}$, $I_i=t_{i}-t_{i-1}$, the probabilities of the six OPs formed by three consecutive ISIs are computed. As discussed in the Introduction, the OPs are defined by the relative values (see the inset of Fig.~\ref{fig:2}): $I_i<I_{i+1}<I_{i+2}$ gives `012', $I_{i+2}<I_{i+1}<I_{i}$ gives `210', etc.  

\section{Results}
%\label{sec3}
 
\begin{figure}[]
\includegraphics[width=6cm]{fig2a}
\includegraphics[width=6cm]{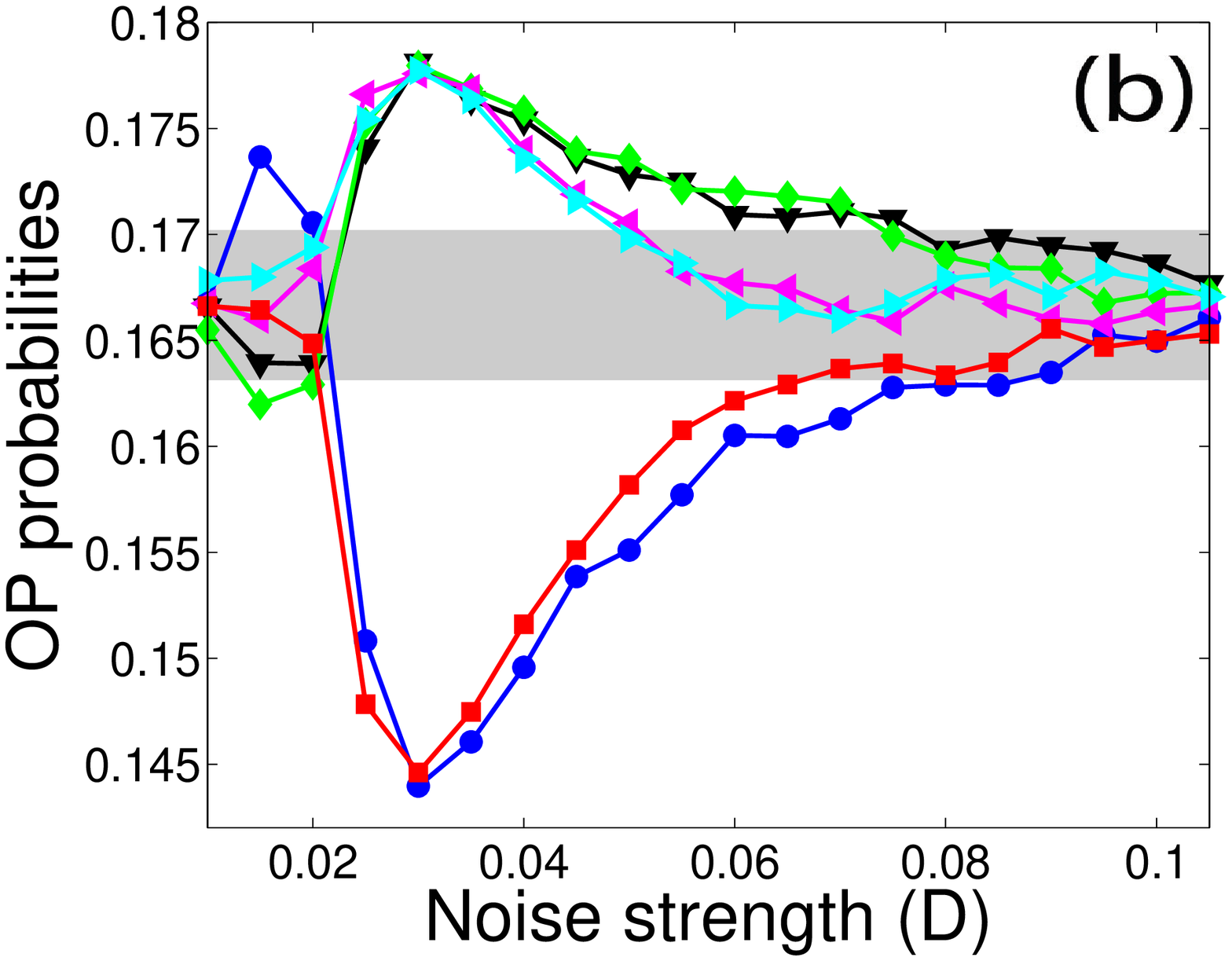}
\includegraphics[width=6cm]{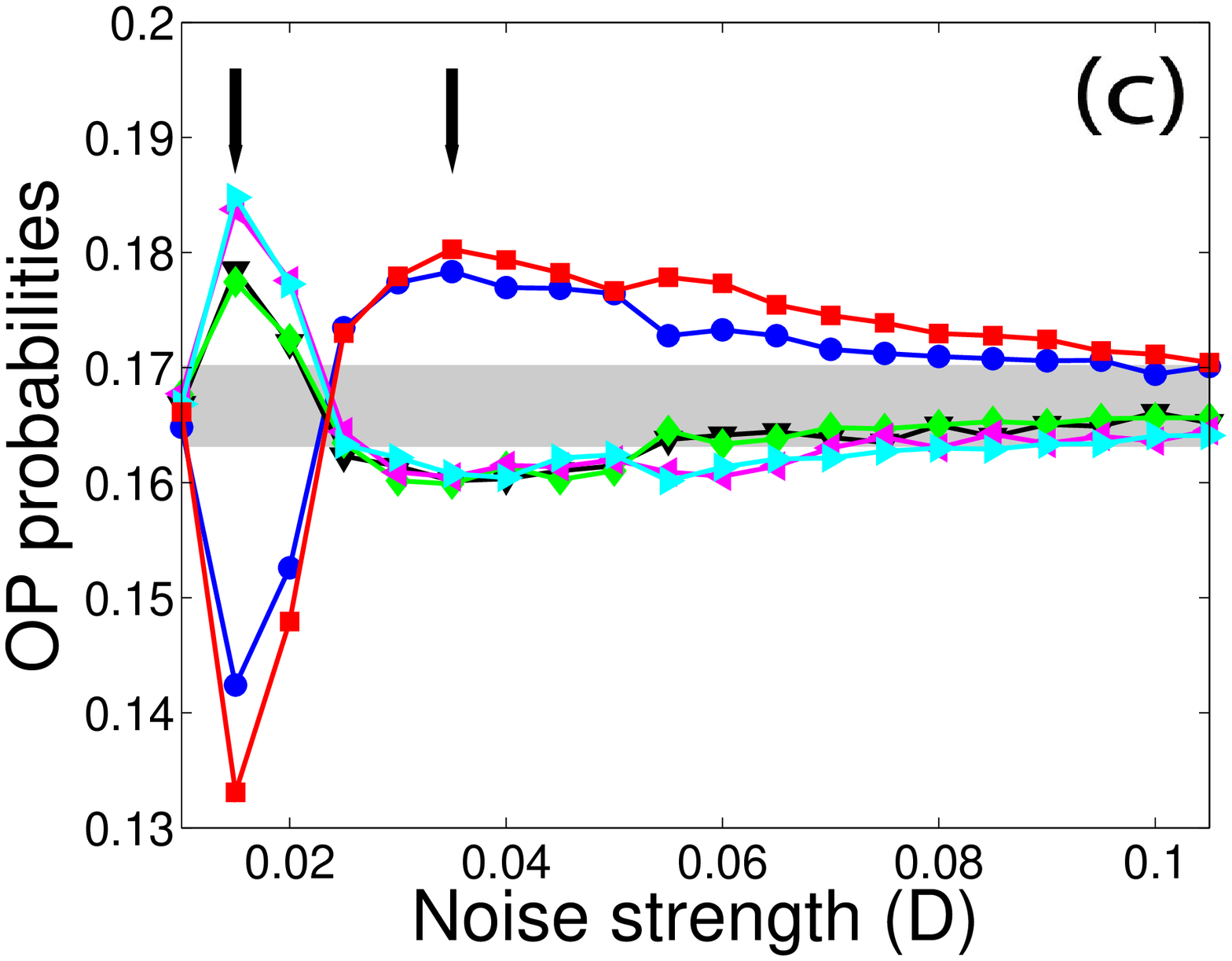}
\caption{(Color online) Probabilities of the six ordinal patterns (OPs) that are defined by the relative length of three consecutive inter-spike-intervals (ISIs) vs. the noise strength. The OPs are schematically shown in the inset. To compute the OP probabilities, time-series with more than 100,000 ISIs were simulated. The parameters are (a) $a_o$=0, (b) $a_o$=0.02, $T=10$, (c) $a_o$=0.02, $T=20$; other parameters are as indicated in the text. In panel (c), the arrows indicate the noise levels used in Figs. 3 and 4.
\label{fig:2}}
\end{figure}

Let us first analyze the ISI sequence generated by the stochastic input only ($a_o=0$). Figure~\ref{fig:2}(a) displays the probabilities of the six OPs as a function of the noise strength, and the grey region indicates the probability region consistent with the uniform distribution \cite{footnote}. We can observe that, within the range of noise strength considered, the six probabilities are in the grey region, and thus, we infer that there are no specific order relations in the ISI sequence. This is due to the fact that the spikes are induced by a fully random process (Gaussian white noise).

Next, we add the weak periodic input, and again plot the OP probabilities vs the noise strength [in Fig.~\ref{fig:2}(b), $T=10$; in Fig.~\ref{fig:2}(c), $T=20$]. We observe a resonance-like phenomenon, in which the probabilities of some patterns lie outside the grey region for certain noise strengths. For example, in Fig.~\ref{fig:2}(b), we note that for $D\sim 0.03$, ``V'' and ``$\Lambda$'' are the preferred patterns; in Fig.~\ref{fig:2}(c), with weak noise ``V'' and ``$\Lambda$'' are preferred, but with higher noise, `012' and `210' are preferred.

\begin{figure}[]
\includegraphics[width=4.2cm]{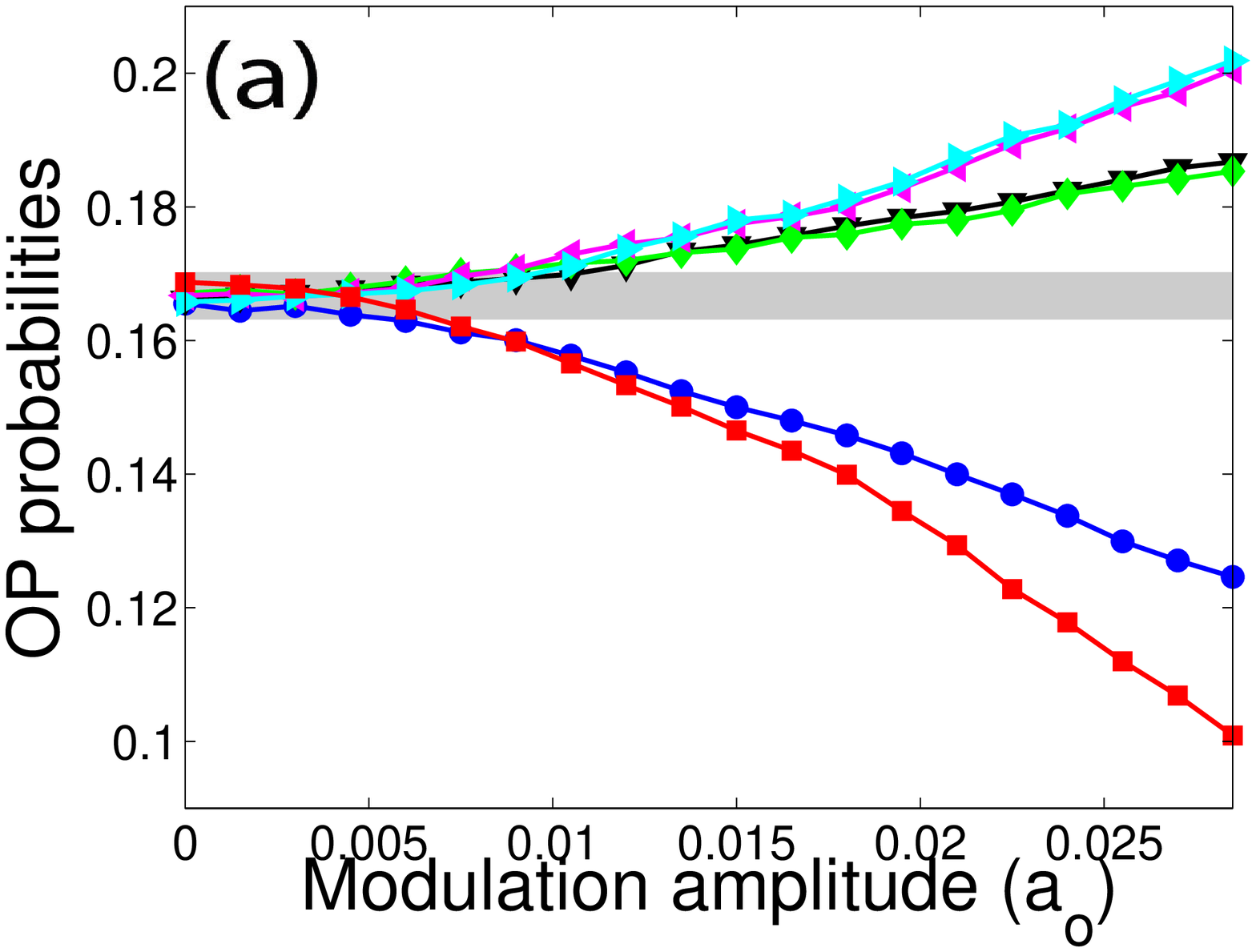}
\includegraphics[width=4.2cm]{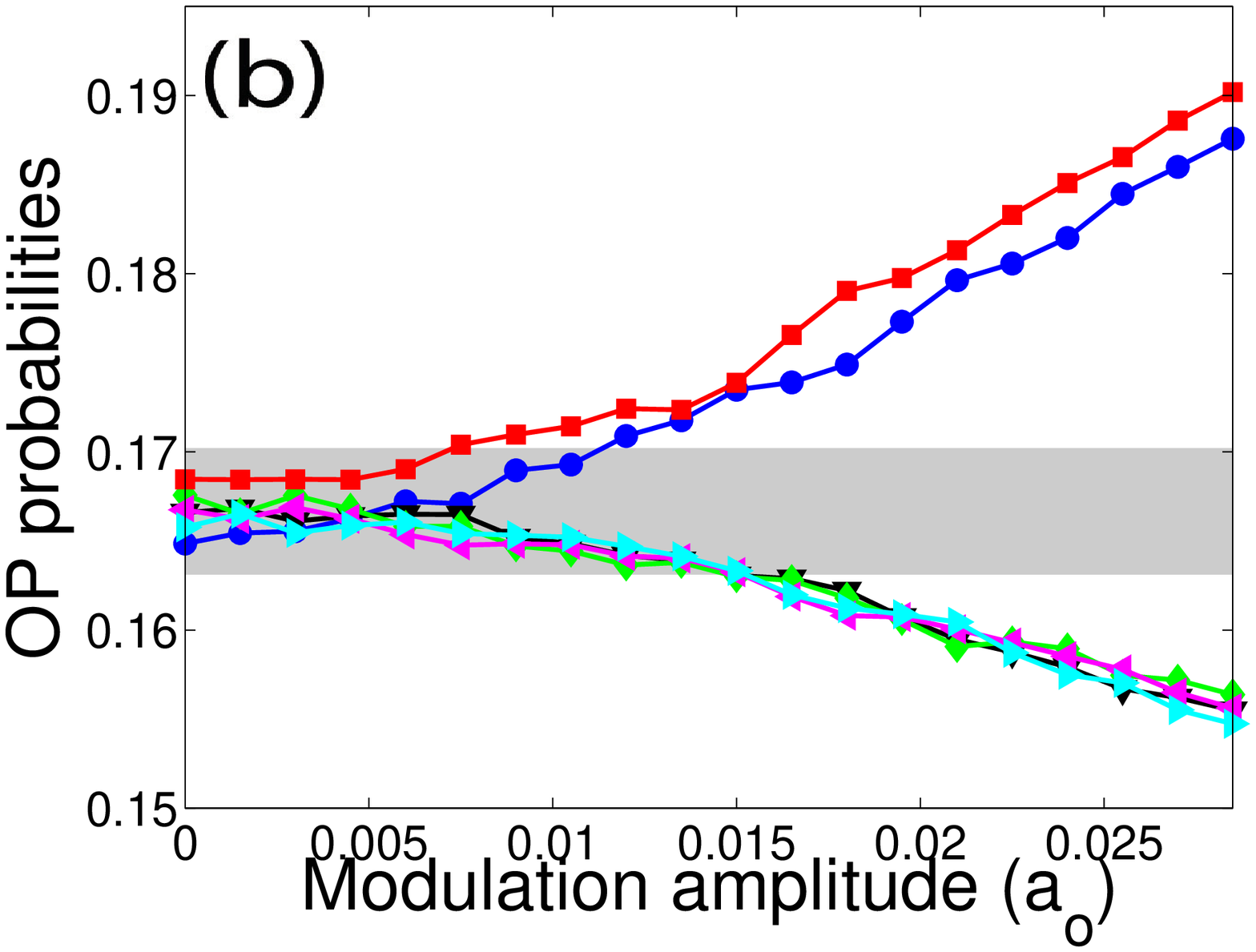}

\caption{(Color online) OP probabilities vs. the amplitude of the input signal. The parameters are $T=20$, (a) $D=0.015$ and (b) $D=0.035$, other parameters as in Fig. 1. \label{fig:3}}
\end{figure}

The effect of the periodic signal gradually increases with its amplitude. This is shown in Fig.~\ref{fig:3} that displays the OP probabilities vs. $a_o$, keeping fixed the period of the signal and the strength of the noise. We consider weak noise [Fig.~\ref{fig:3}(a)] and stronger noise [Fig.~\ref{fig:3}(b)], which induce different ISI order relations [as indicated with arrows in Fig.~\ref{fig:2}(c)]. We observe that, in both cases, as $a_o$ increases, the OP probabilities gradually leave the grey region, revealing that order relations gradually emerge in the ISI sequence. We note that, within the range of values considered here (the input is subthreshold), $a_{o}$ does not change the preferred OPs. 

\begin{figure}[]
\includegraphics[width=4.2cm]{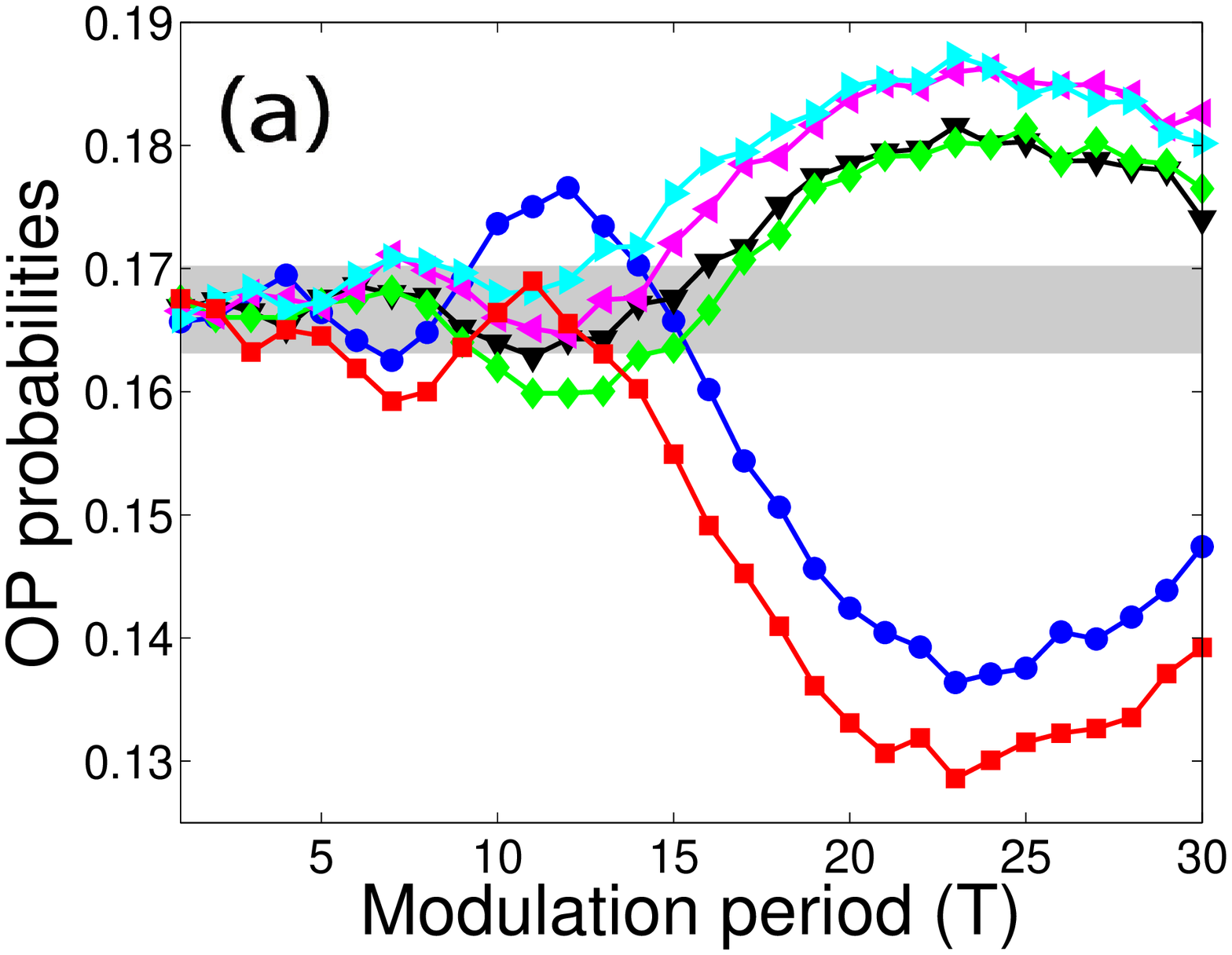}
\includegraphics[width=4.2cm]{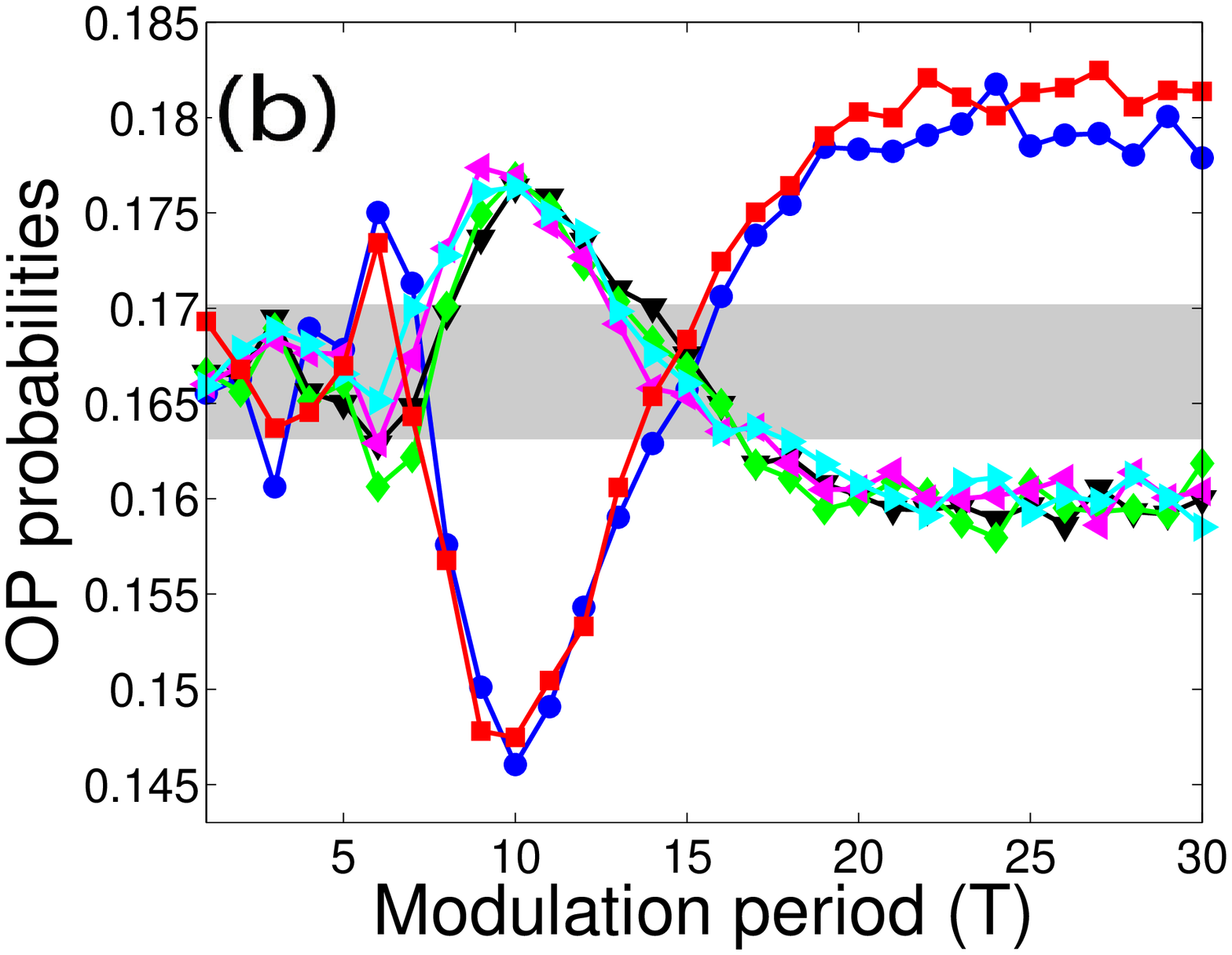}
\caption{(Color online) OP probabilities vs. the period of the input signal. The parameters are $a_o$=0.02, (a) $D=0.015$ and (b) $D=0.035$, other parameters as in Fig. 1.
\label{fig:4}}
\end{figure}

In order to investigate the role of the period of the input signal, in Fig.~\ref{fig:4} we display the OP probabilities vs. $T$. We consider weak and stronger noise (the same levels as in Fig.~\ref{fig:3}). We note that when the input signal is fast, the OP probabilities are inside the gray region, but for slower input, they lie outside. We also note that the preferred patterns depend on both, $T$ and $D$, and there is a resonant-like effect in the form of enhanced probability of particular OPs for specific values of $T$ and $D$. For example, for $D = 0.035$ [Fig.~\ref{fig:4}(b)] patterns `012' and `210'  are preferred for $T\sim 6$, but they are unlikely to occur for $T\sim 10$.

%For example, for $T=10$, $(012)$ and $(210)$ are the more probable patterns for low noise [Fig. 3(a)], but they become less probable with higher noise [Fig. 3(b)]. For longer periods, the opposite ordering occurs. For example, for $T=20$, $(012)$ and $(210)$ are the less patterns in panel (a), and the more probable patterns in panel (b).

% 
% \begin{figure}[]
% \includegraphics[width=6cm]{fig2b1}
% \includegraphics[width=6cm]{fig2b2}
% 
% \caption{(Color online) OP probabilities vs. the amplitude of the sinusoidal input signal. The parameters are $D=0.035$, (a) $T=10$ and (b) $T=20$, other parameters as in Fig. 1. \label{fig:2}}
% \end{figure}

\begin{figure}[]
\includegraphics[width=4.2cm]{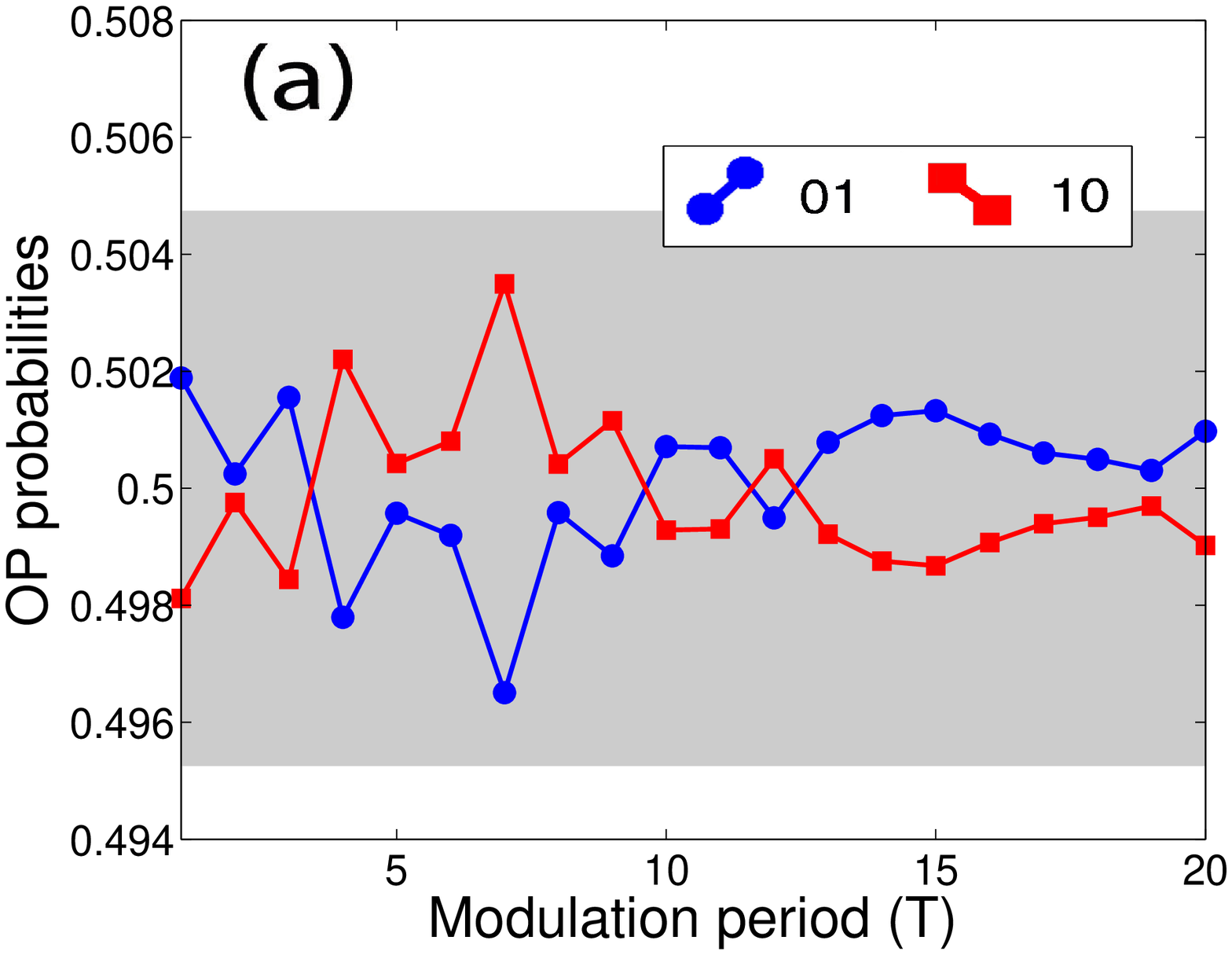}
\includegraphics[width=4.2cm]{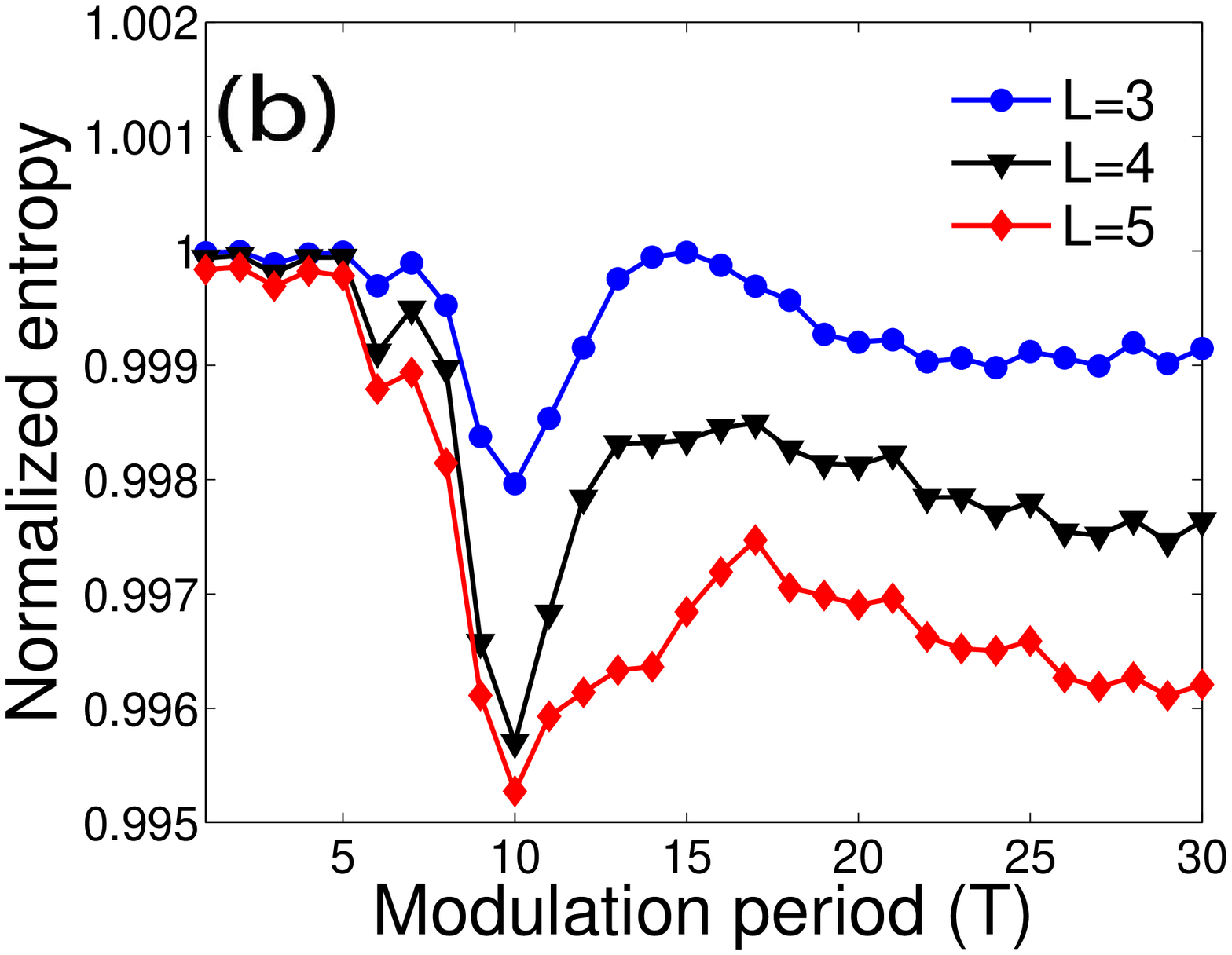}
\caption{(Color online) (a) Probabilities of patterns `01' and `10' vs. the period of the input signal. (b) Permutation entropy vs. $T$ for OPs of length $L=$3, 4, and 5. In panels (a) and in (b) the parameters are as in Fig. 4(b).
\label{fig:5}}
\end{figure}

To explore the length of temporal ordering, we show in Fig.~\ref{fig:5}(a), for the same parameters as Fig.~\ref{fig:4}(b), the probabilities of OPs of length $L=2$: $I_i<I_{i+1}$ gives pattern `01' and $I_i>I_{i+1}$ gives `10' \cite{footnote2}. We observe that they are in the grey area, which indicates that there is no temporal order in the ISI sequence. However, the probabilities of $L=3$ OPs revealed the presence of  patterns with favored occurrence, as it was shown in Fig.~\ref{fig:4}(b). Therefore, we conclude that, in order to uncover temporal ordering, the ISI sequence has to be analyzed with OPs of appropriate length. To explore the effect of longer OPs, it is unpractical to display the probabilities, $p_i$, of the $L!$ OPs, because there are 24 $L=4$ OPs and 120 $L=5$ OPs. Therefore, in Fig.~\ref{fig:5}(b) we plot the permutation entropy \cite{bp_prl_2002,cita2}, $H$, computed with patterns of length $L$=3, 4, and 5 vs. the period of the input signal. In this way we uncover a clear transition as $T$ increases: for $T<5$, $H\sim 1$, while for longer $T$, $H$ tends to decrease, but non-monotonically, i.e., there are values of $T$ for which $H$ is minimum, indicating the existence of more probable patterns and thus, temporal ordering in the ISI sequence. We also note that, while for $T<5$ $H\sim 1$ for $L=$3-5, for $T>5$, the permutation entropy decreases with $L$, indicating the longer range of temporal ordering. 

Analyzing the ISI sequence with longer OPs is computationally very expensive as the large number of possible OPs requires extremely long time series in order to compute the probabilities with good statistics. This is shown in Fig. 6, that displays the OP probabilities vs. the length of the dataset, $M$. We see that, with a periodic input signal [panels 6(a) and 6(b)], the OP probabilities are outside the grey region, if $M$ is large enough. Moreover, in panel 6(b), ``clusters'' of OPs with similar probabilities are seen, only if $M>>10^3$ (similar clustering was reported in \cite{Aragoneses:2014sr}). In contrast, without periodic input [panel 6(c)] the probabilities are inside the grey region and no clustering is seen, even for large $M$.

\begin{figure}[]
\includegraphics[width=6cm]{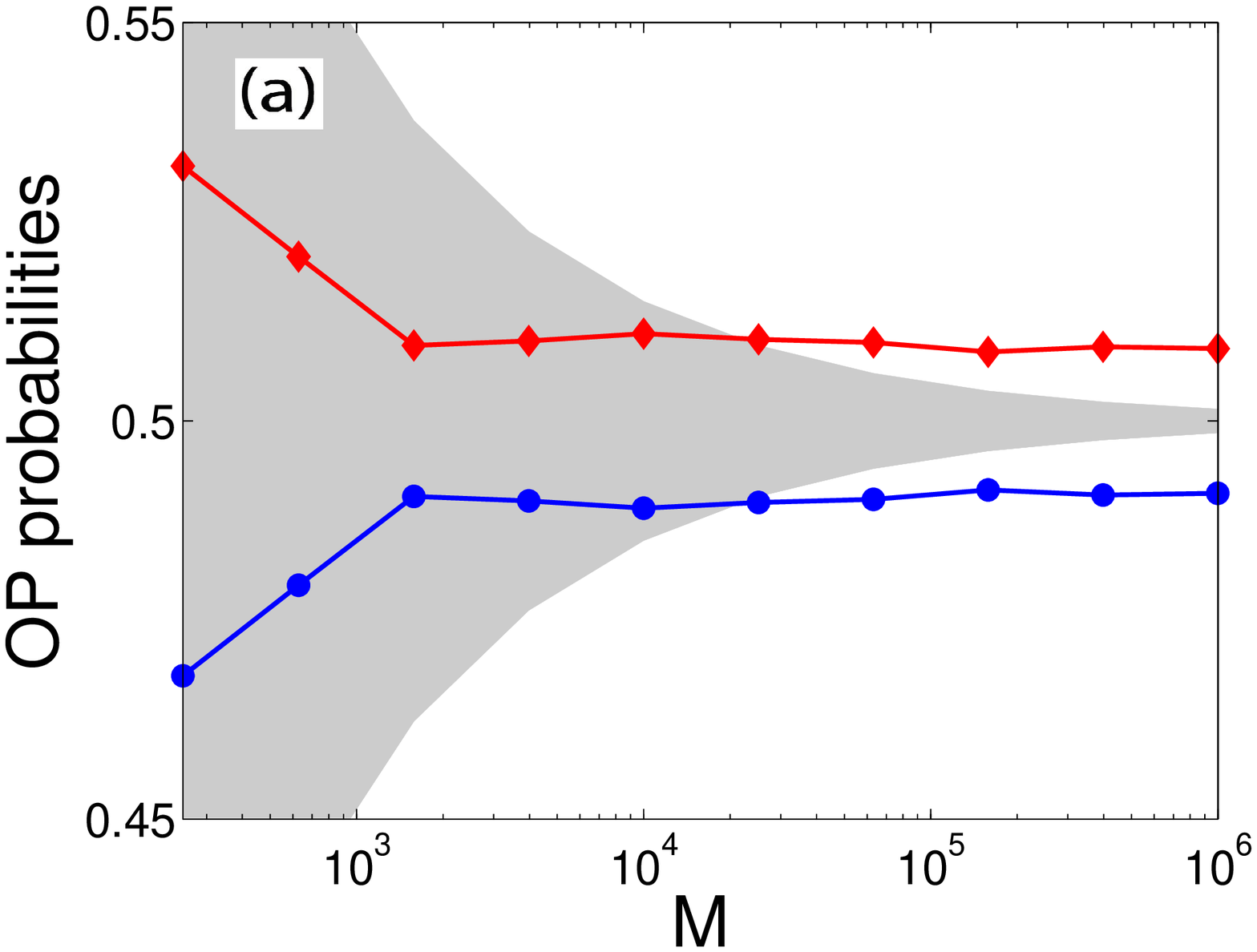}
\includegraphics[width=6cm]{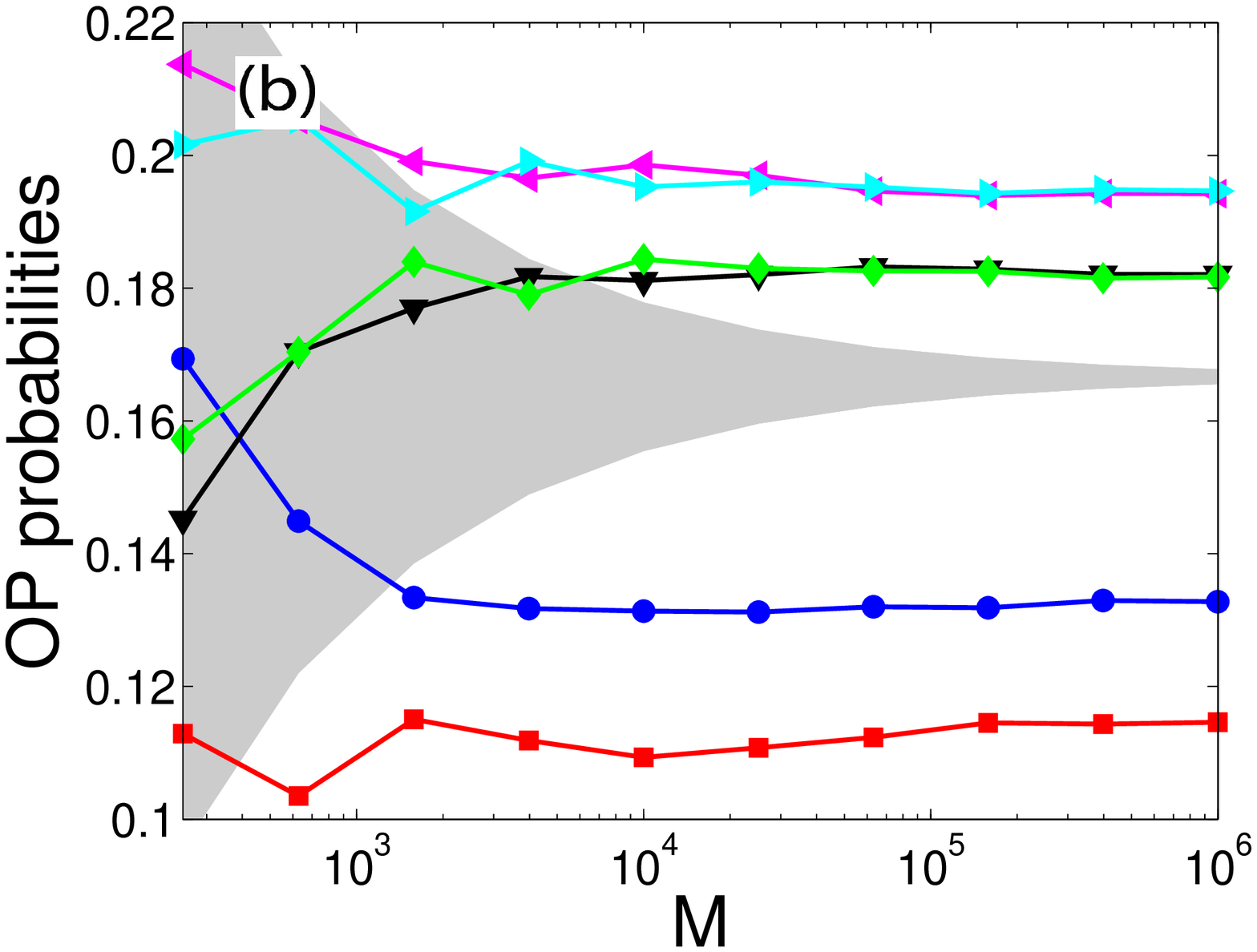}
\includegraphics[width=6cm]{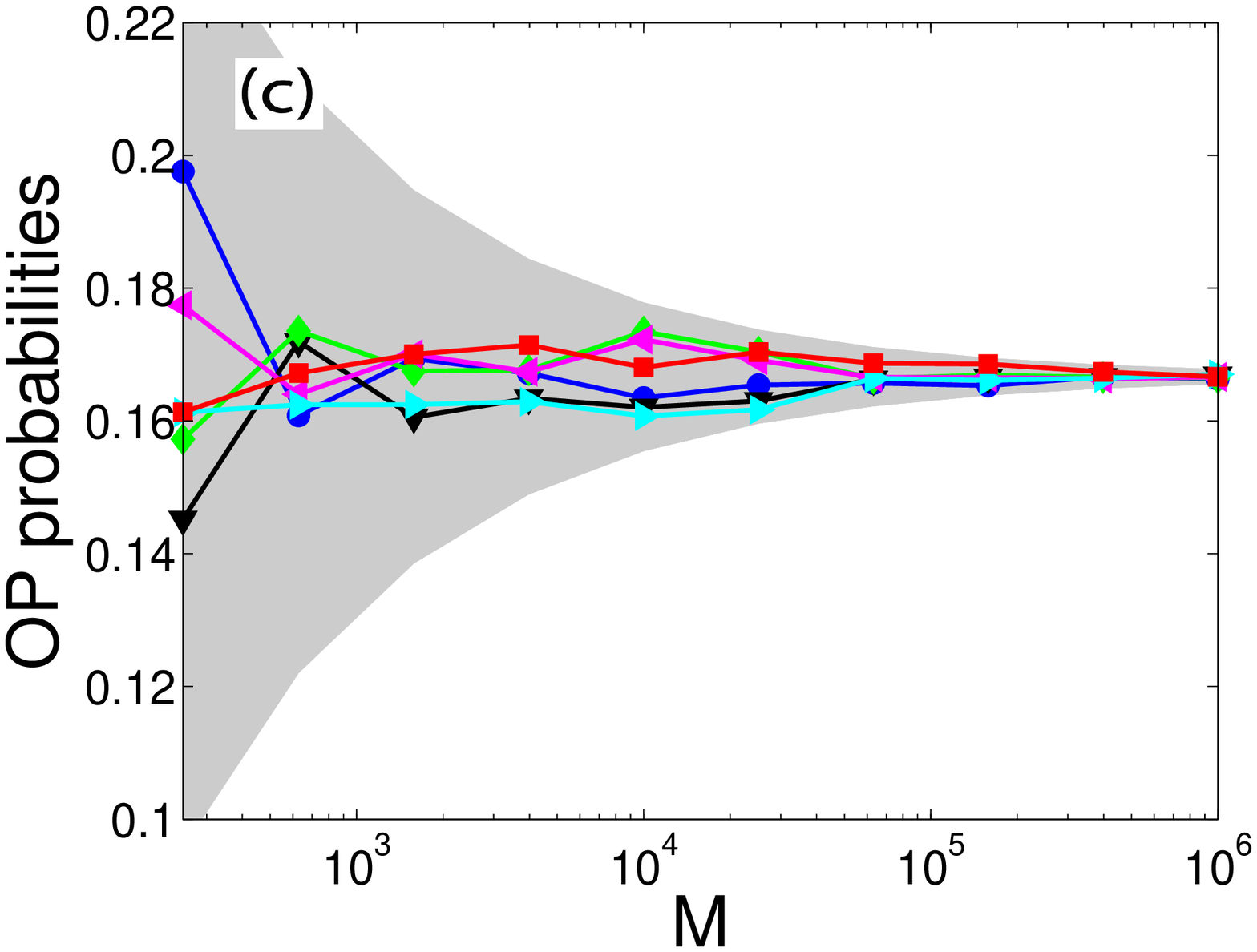}
\caption{(Color online) (a) Probabilities of patterns `01' and `10' vs. the number, $M$, of inter-spike intervals in logarithmic scale.  (b) Probabilities of the 6 $L=3$ OPs vs $M$. For both, (a) and (b), the parameters are $a_{o}=0.025$, $D=0.015$ and $T=20$. 
(c) Same as panel (b) but with $a_{o}=0$.
Figs. 2-5 were done with $M=10^{5}$. As it can be appreciated mainly in panel (a), this might not be enough to detect significant differences among OP probabilities.}
\label{fig:6}
\end{figure}

Interestingly, the behavior of the OP probabilities seen in Fig.~\ref{fig:3}(a) resembles that found experimentally in a modulated semiconductor laser that emits feedback-induced optical spikes \cite{Aragoneses:2014sr}. As shown in Fig. 4(a) in \cite{Aragoneses:2014sr}, when the modulation amplitude increases there is a transition to a dynamical state in which some OP probabilities are outside the grey region, and, remarkable, the OP probabilities are organized in the same ``clusters'', and with the same hierarchy (the same ordering of the OP probabilities) as observed in Fig.~\ref{fig:3}(a) here. This qualitative similarity can be due to a generic behavior of excitable systems, that can be described by circle maps \cite{piro}. As shown in \cite{Aragoneses:2014sr}, a modified circle map qualitatively explains the behavior of the OP probabilities computed from the laser data, and it has been shown to also explain serial correlations in empirical ISI data \cite{neiman_pre}. This suggests that similar behavior can be observed in other excitable systems.

\section{Comparison with mean-ISI and correlation analysis}

\begin{figure}[htbp]
\centering
 \includegraphics[width=2.8 cm]{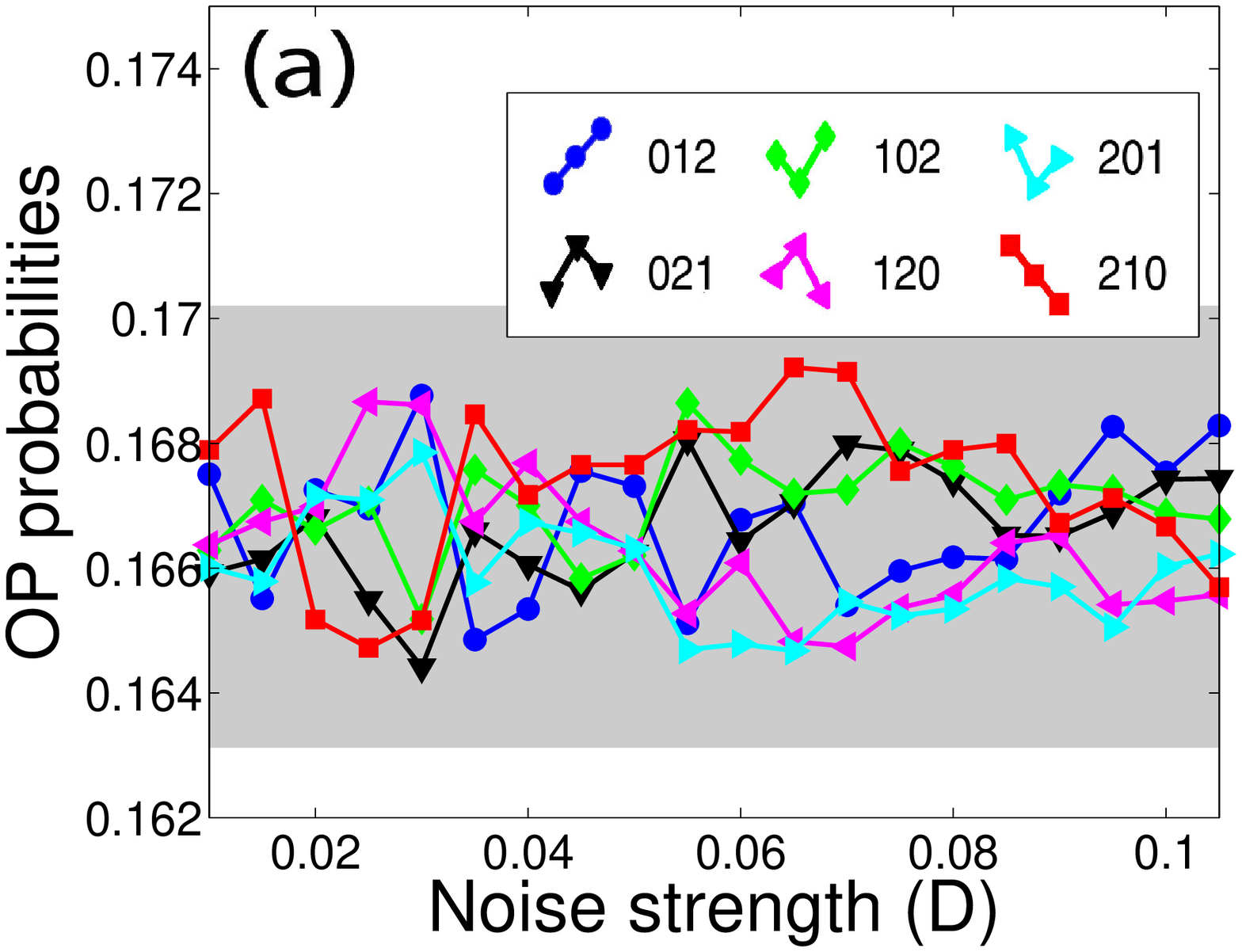}
 \includegraphics[width=2.8 cm]{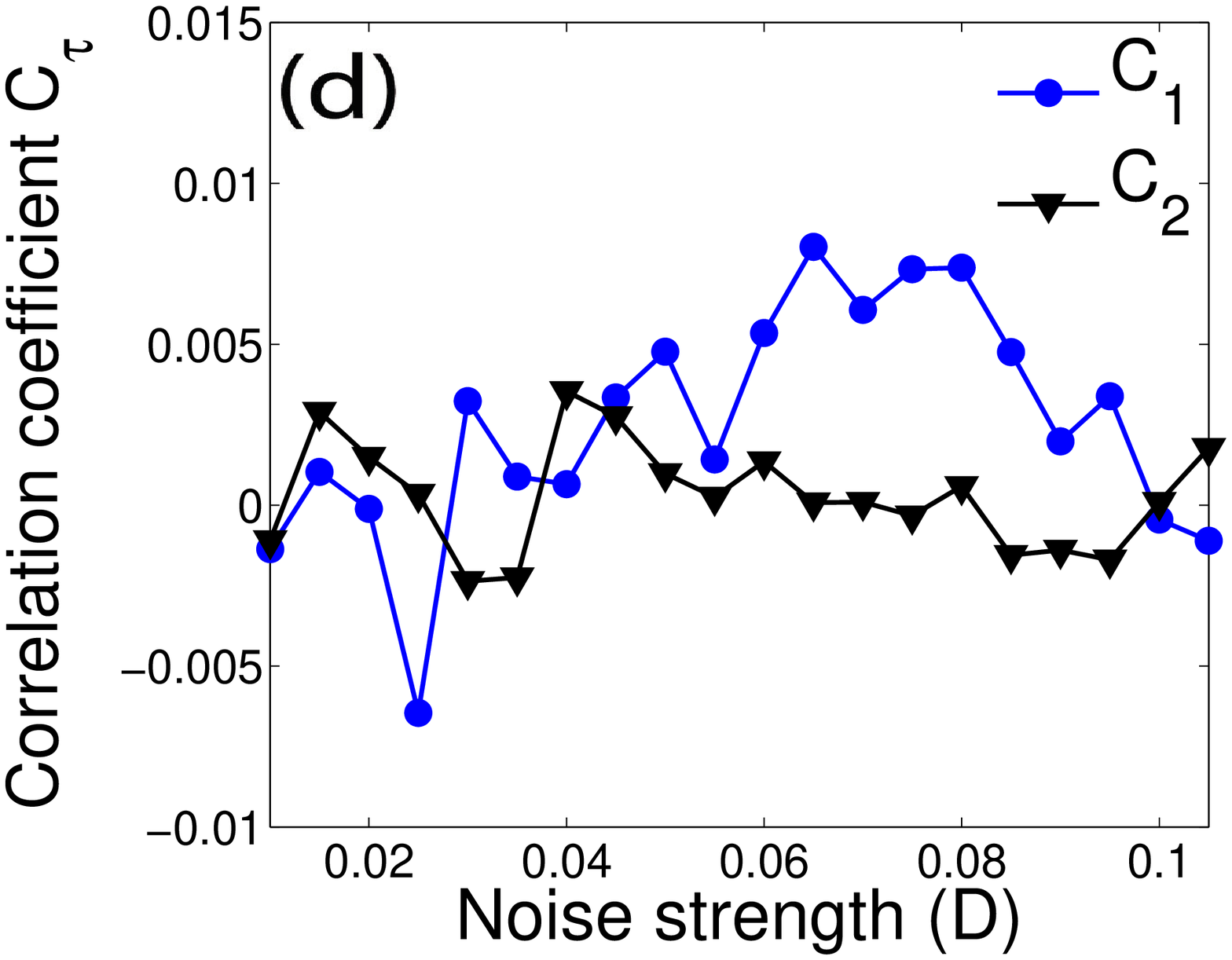}
 \includegraphics[width=2.8 cm]{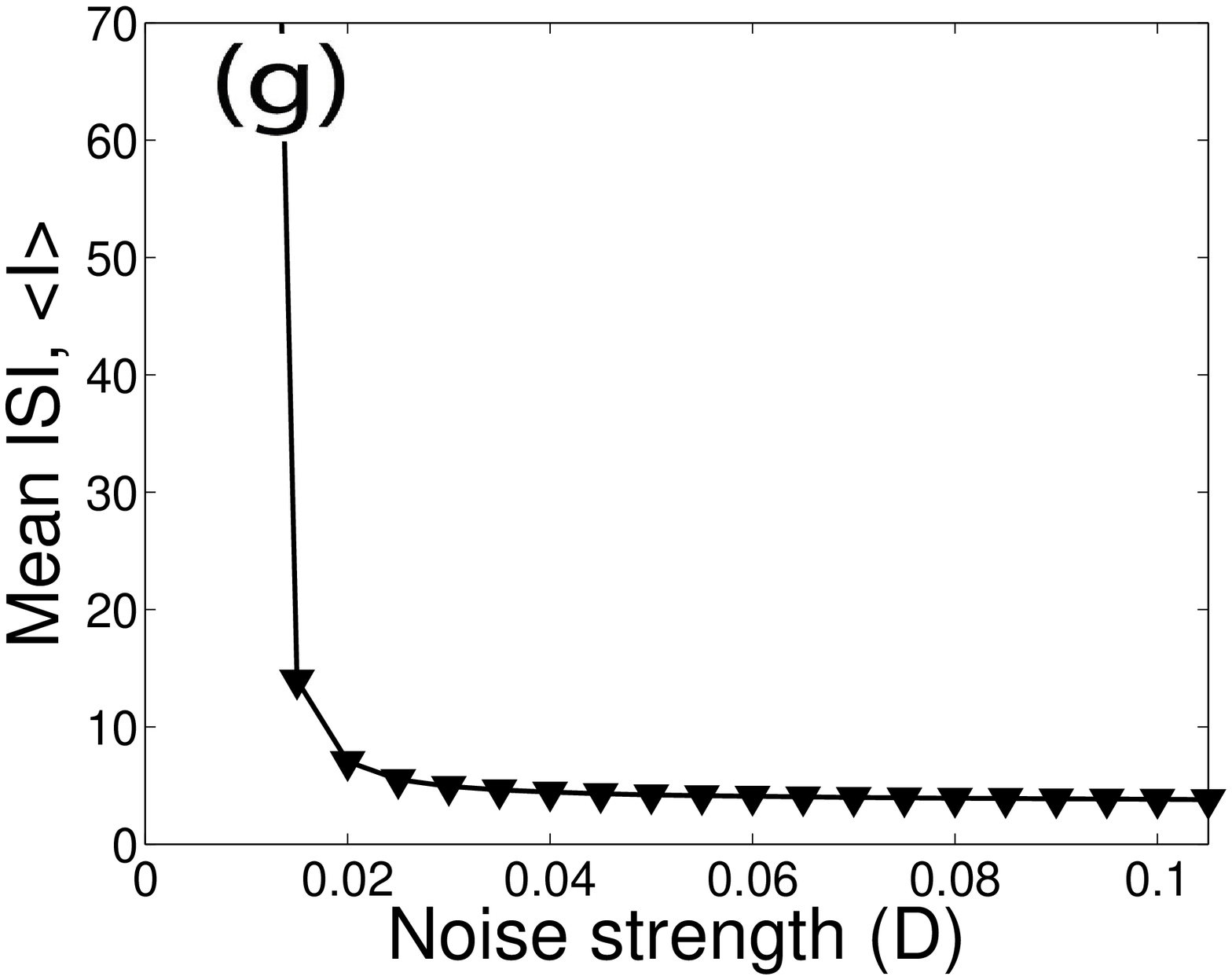}\hfill
 \includegraphics[width=2.8 cm]{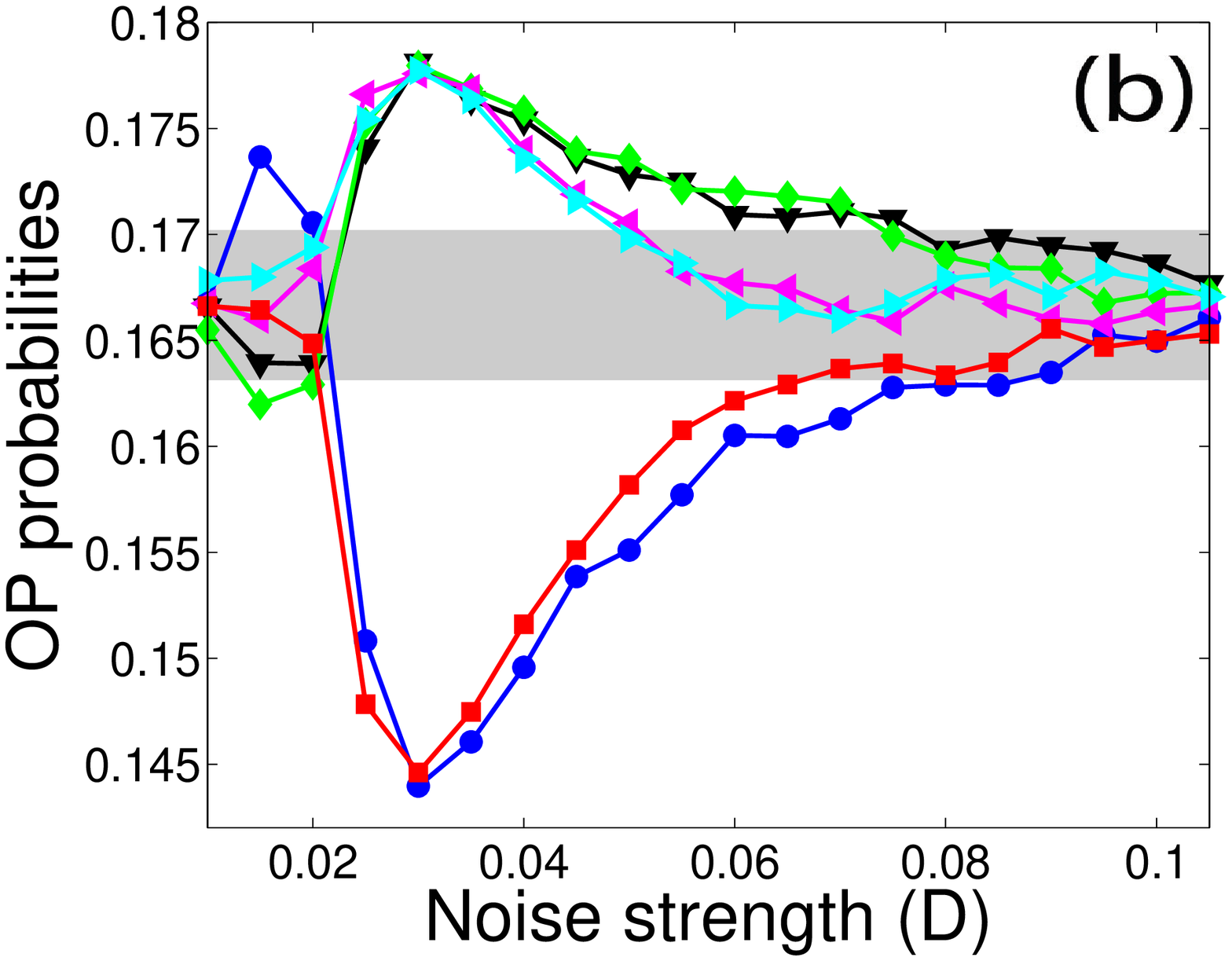}
 \includegraphics[width=2.8 cm]{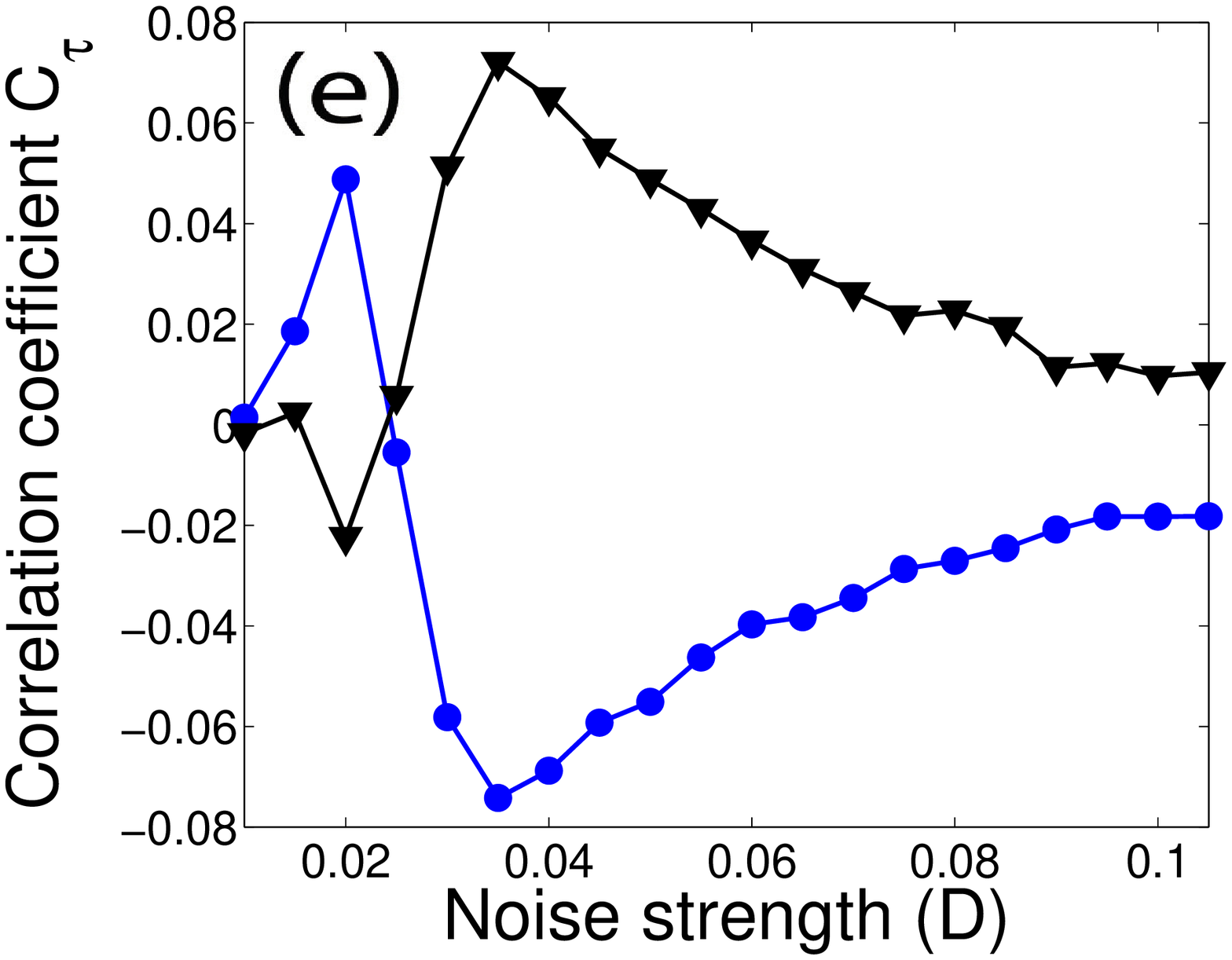}
 \includegraphics[width=2.8 cm]{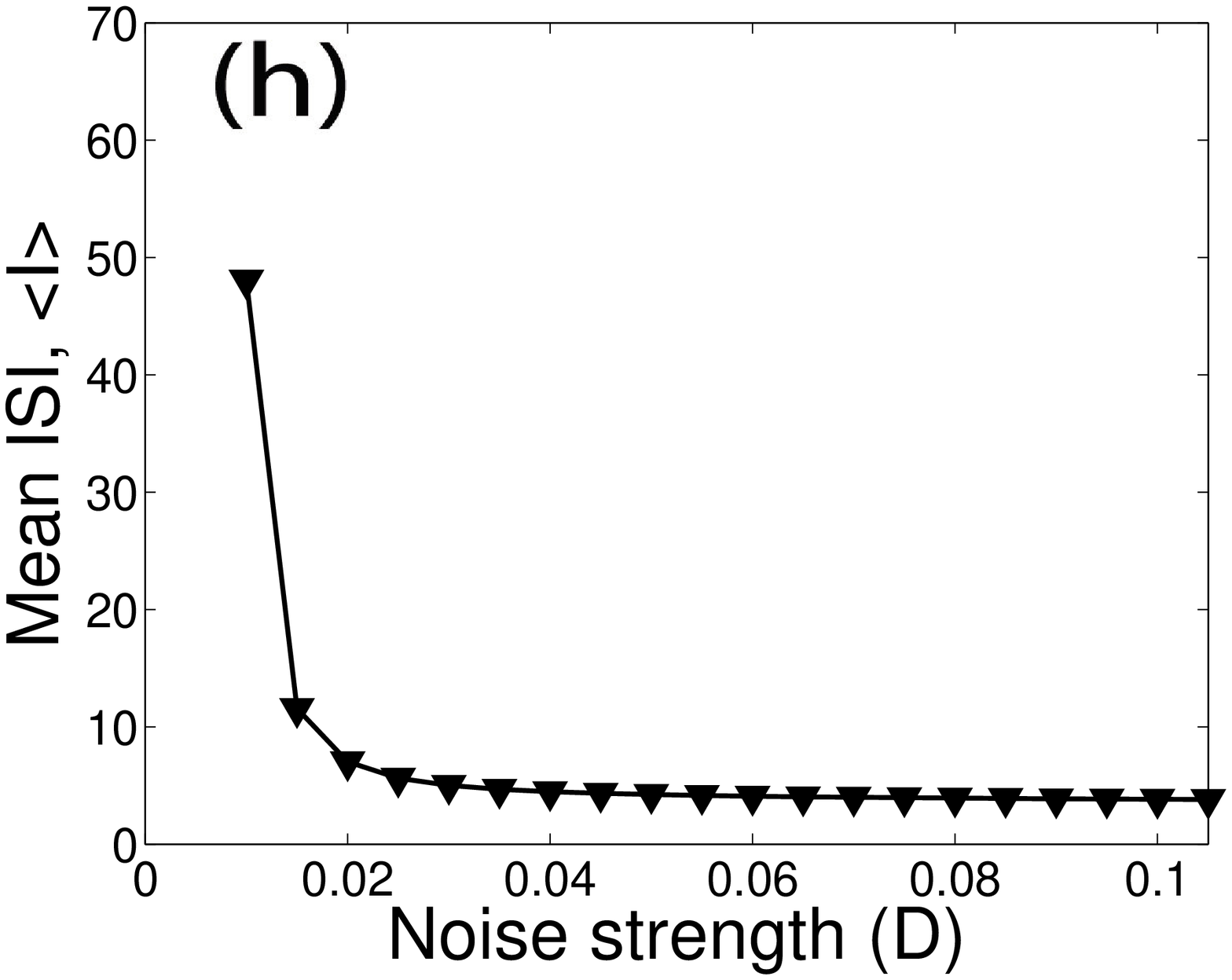}\hfill
 \includegraphics[width=2.8 cm]{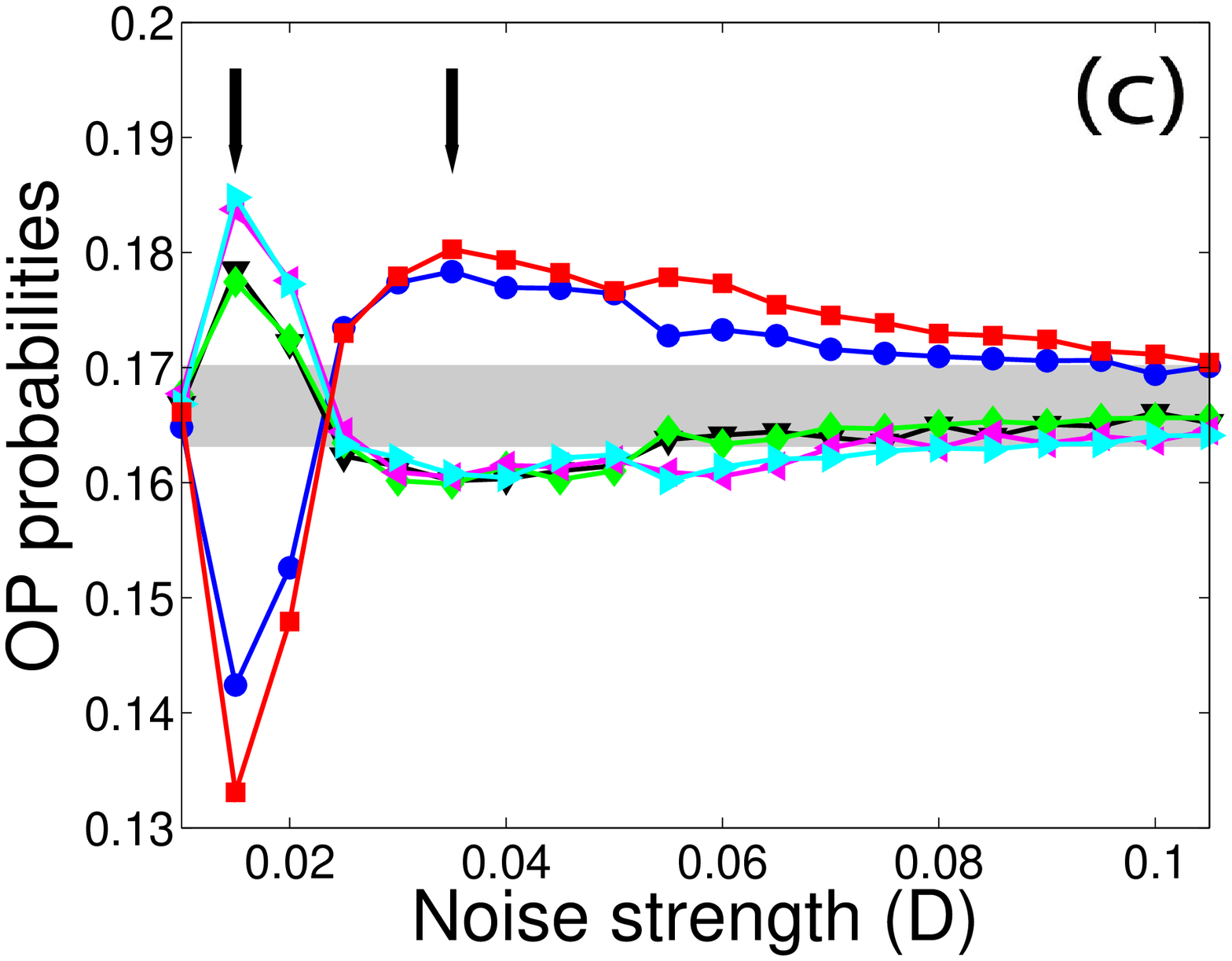}
 \includegraphics[width=2.8 cm]{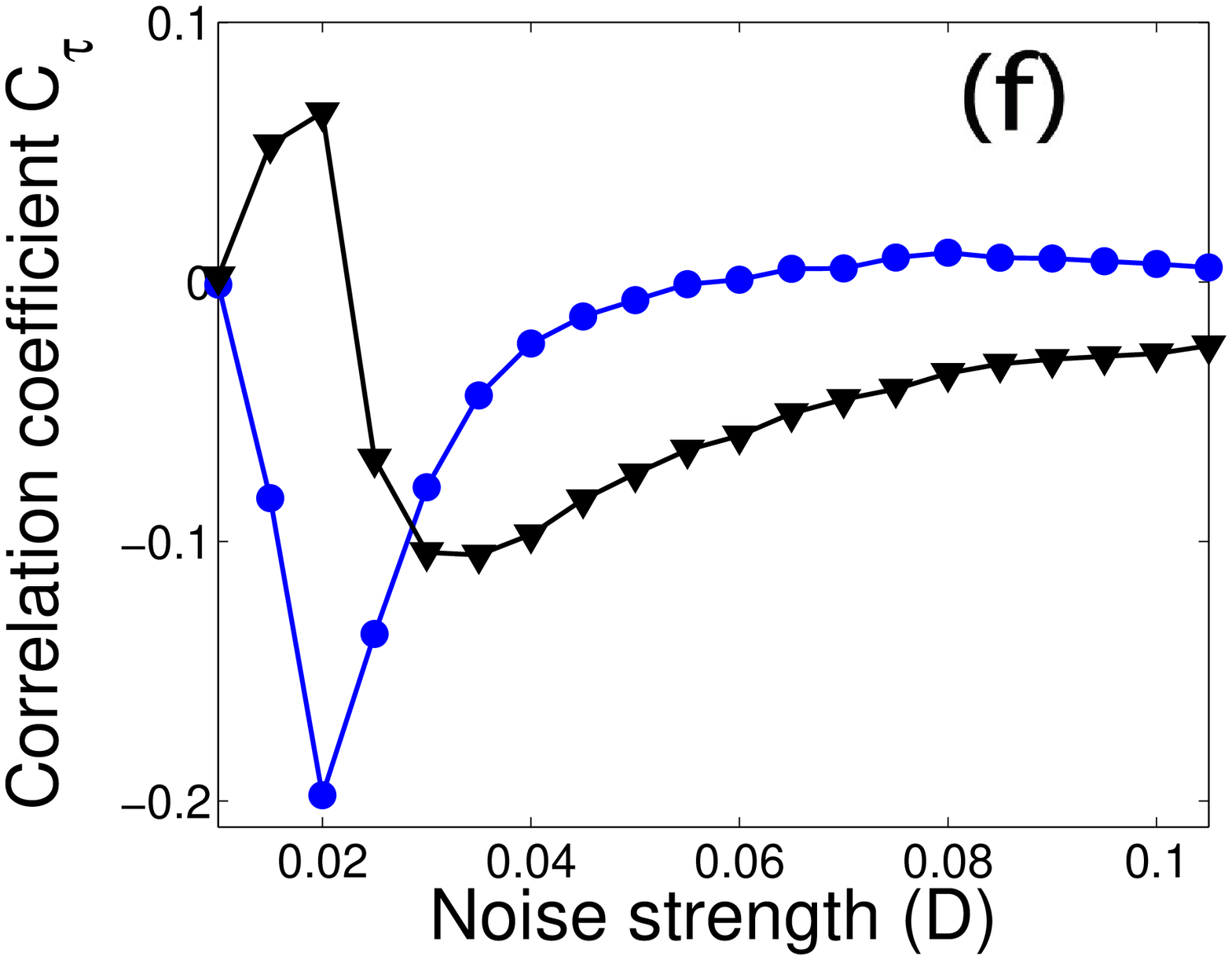}
 \includegraphics[width=2.8 cm]{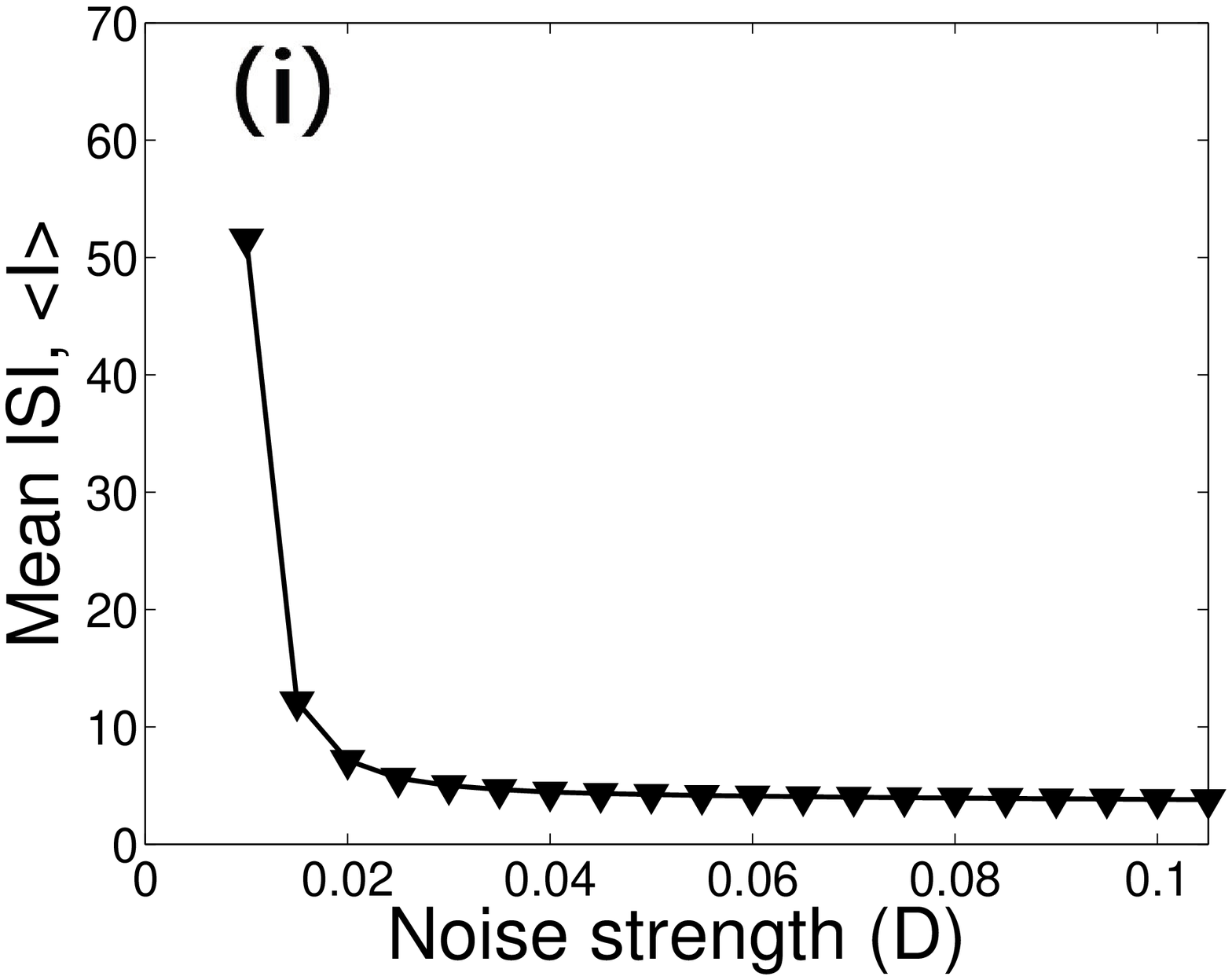}
% \subfloat{\includegraphics[width=0.15\textwidth]{fig.s1a.eps}}
% \subfloat{\includegraphics[width=0.15\textwidth]{fig.s1d.eps}}
% \subfloat{\includegraphics[width=0.15\textwidth]{fig.s1g.eps}}\hfill
% \subfloat{\includegraphics[width=0.15\textwidth]{fig.s1b.eps}}
% \subfloat{\includegraphics[width=0.15\textwidth]{fig.s1e.eps}}
% \subfloat{\includegraphics[width=0.15\textwidth]{fig.s1h.eps}}\hfill
% \subfloat{\includegraphics[width=0.15\textwidth]{fig.s1c.eps}}
% \subfloat{\includegraphics[width=0.15\textwidth]{fig.s1f.eps}}
% \subfloat{\includegraphics[width=0.15\textwidth]{fig.s1i.eps}}
\caption{(Color online) Left column: OPs probabilities; central column: correlation coefficients and right column: mean inter-spike-interval. The  parameters are as in Fig. 2:  (a) $a_o$=0, (b) $a_o$=0.02, $T=10$, (c) $a_o$=0.02, $T=20$.
\label{fig:7}}
\end{figure}

Since both, the noise strength, $D$, and the period of the input signal, $T$, modify the neuron's spike rate, one could expect that the underlying reason for the variation of the OP probabilities with $D$ and $T$ is related to the spike rate variation. One could also wonder if these changes are also captured by correlation analysis.

\begin{figure}[htbp]
\centering

 \includegraphics[width=2.8 cm]{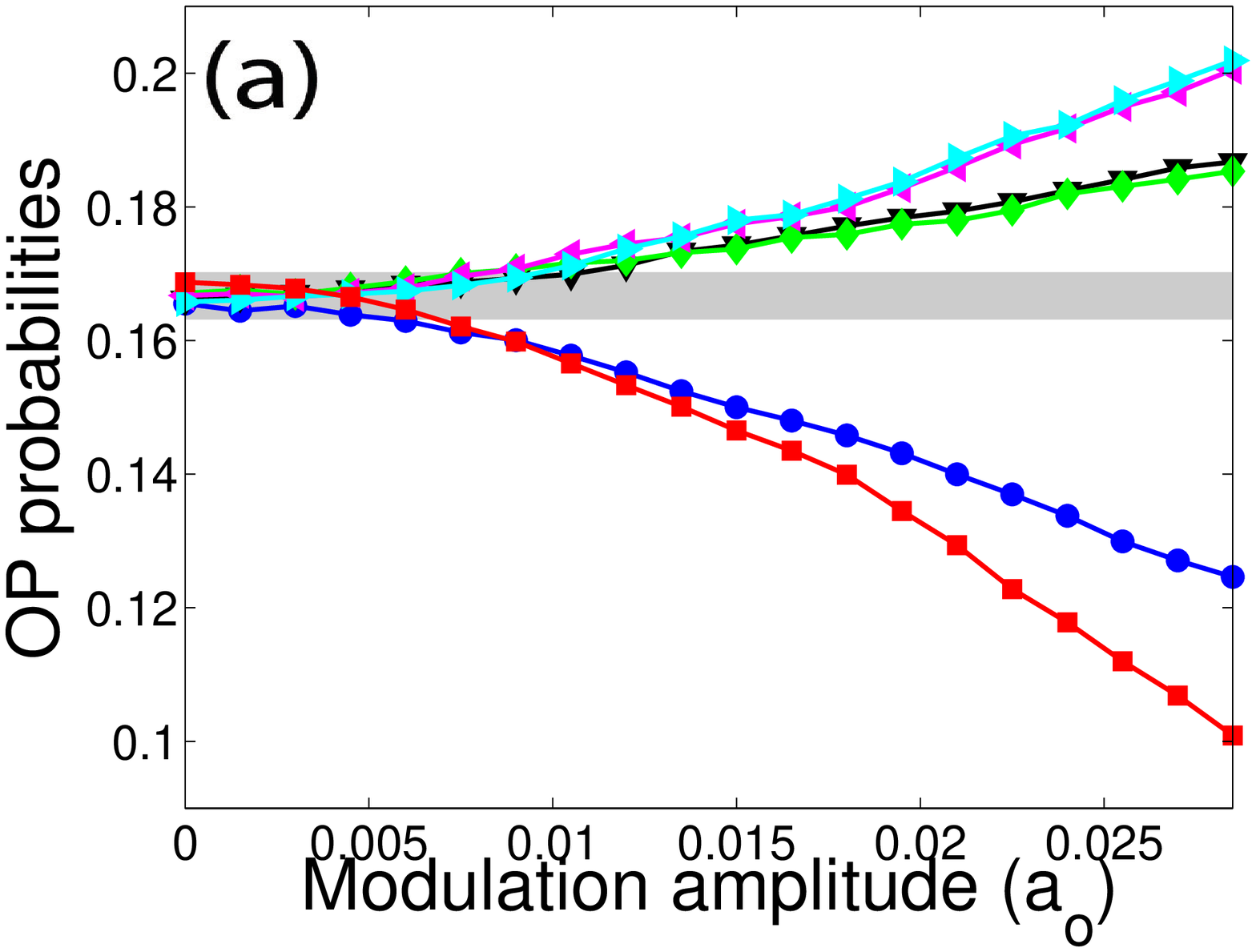}
 \includegraphics[width=2.8 cm]{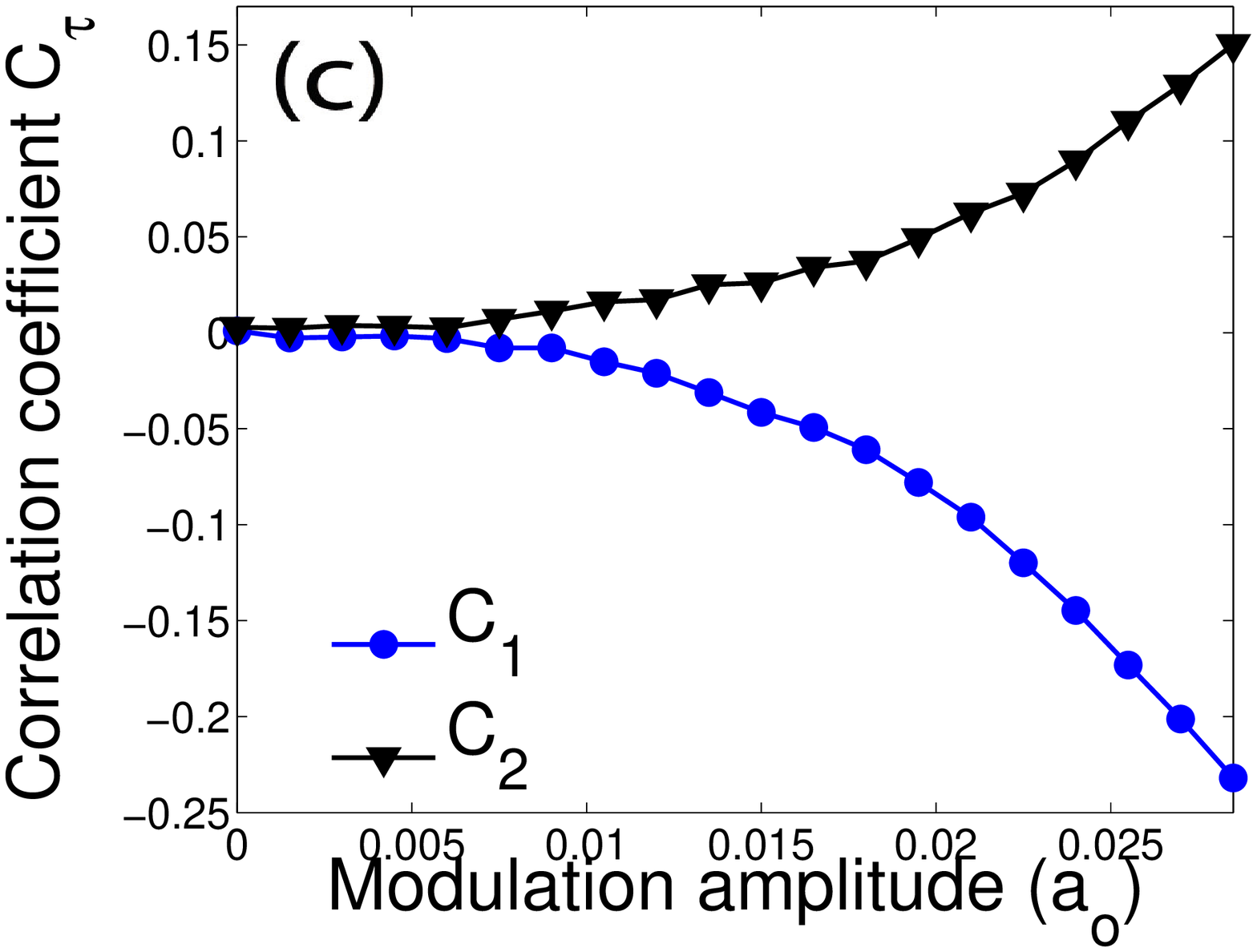}
 \includegraphics[width=2.8 cm]{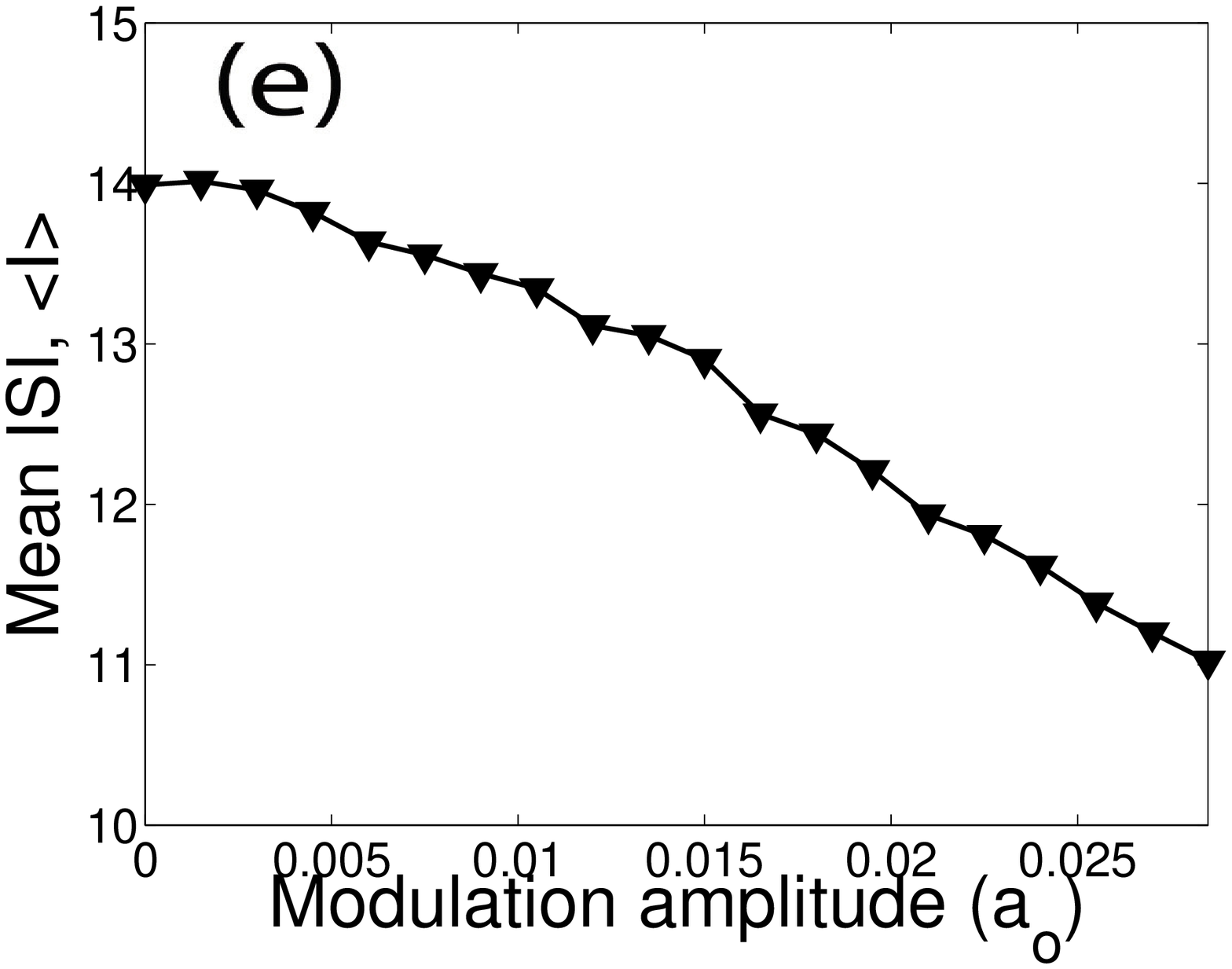}\hfill
 \includegraphics[width=2.8 cm]{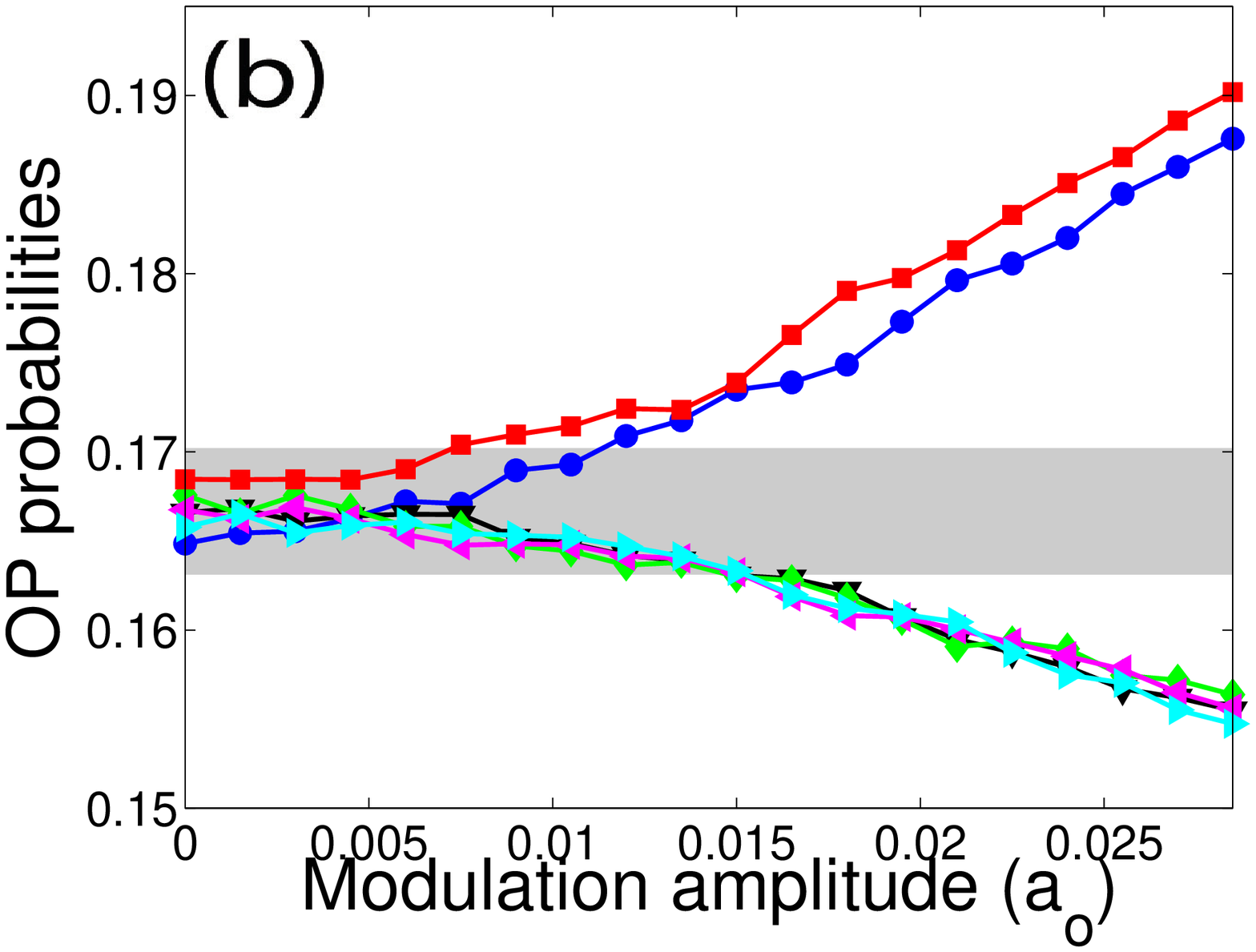}
 \includegraphics[width=2.8 cm]{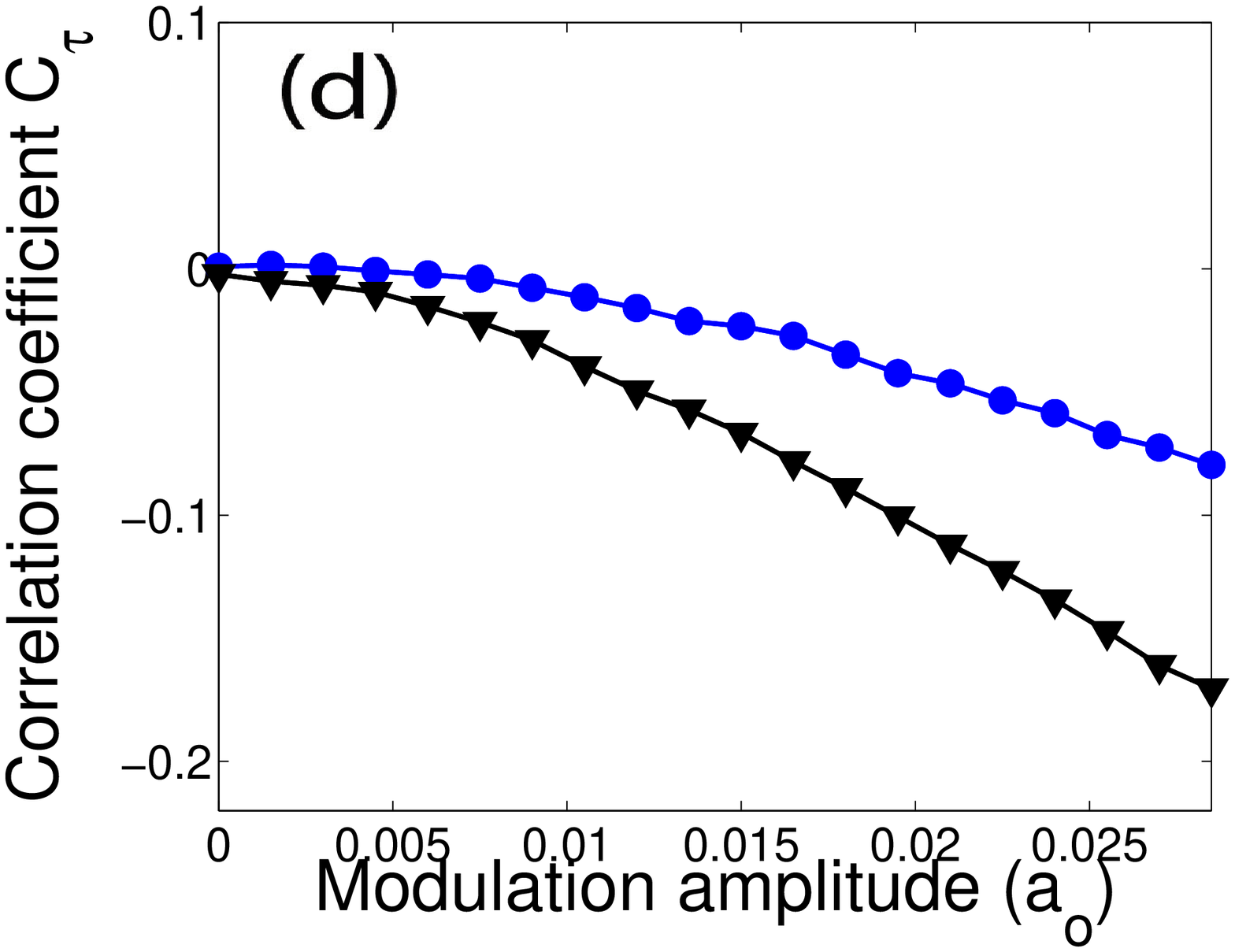}
 \includegraphics[width=2.8 cm]{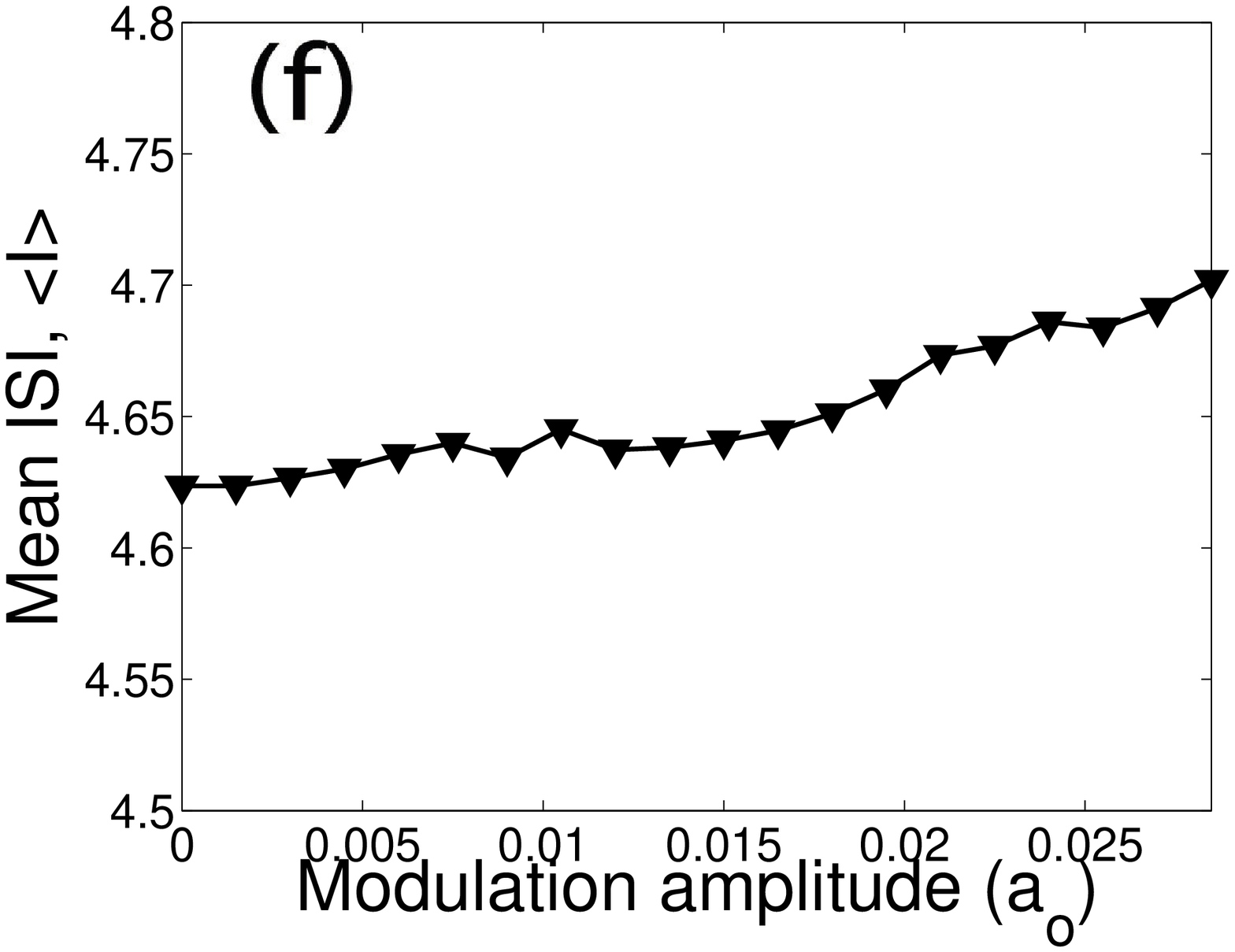}
% \subfloat{\includegraphics[width=0.15\textwidth]{fig.s2a.eps}}
% \subfloat{\includegraphics[width=0.15\textwidth]{fig.s2c.eps}}
% \subfloat{\includegraphics[width=0.15\textwidth]{fig.s2e.eps}}\hfill
% \subfloat{\includegraphics[width=0.15\textwidth]{fig.s2b.eps}}
% \subfloat{\includegraphics[width=0.15\textwidth]{fig.s2d.eps}}
% \subfloat{\includegraphics[width=0.15\textwidth]{fig.s2f.eps}}
\caption{(Color online) Left column: OPs probabilities; central column: correlation coefficients and right column: mean inter-spike-interval. The  parameters are as in Fig. 3: $T=20$, (a) $D=0.015$ and (b) $D=0.035$.
\label{fig:8}}
\end{figure}  

\begin{figure}[htbp]
 \centering
 
 \includegraphics[width=2.8 cm]{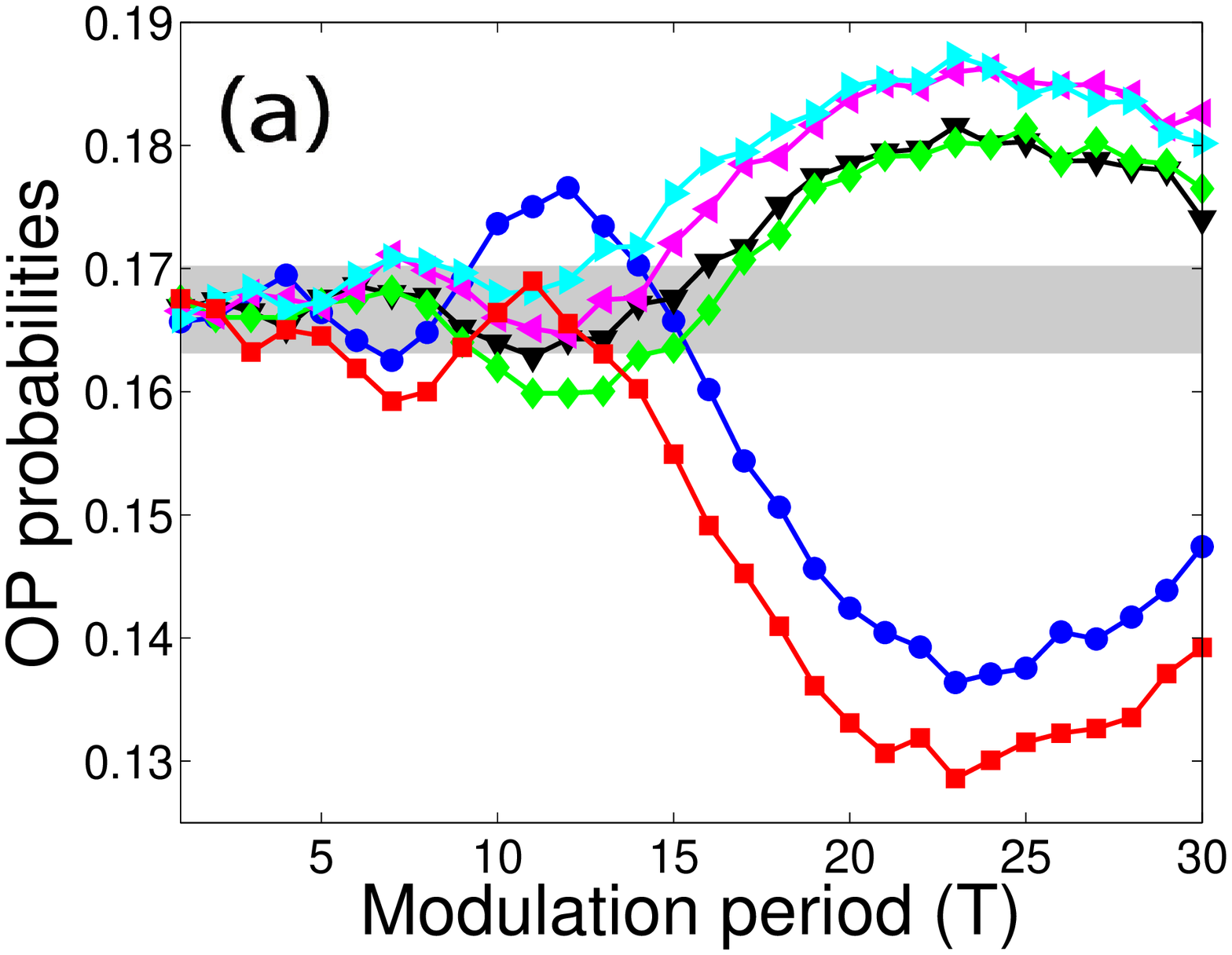}
 \includegraphics[width=2.8 cm]{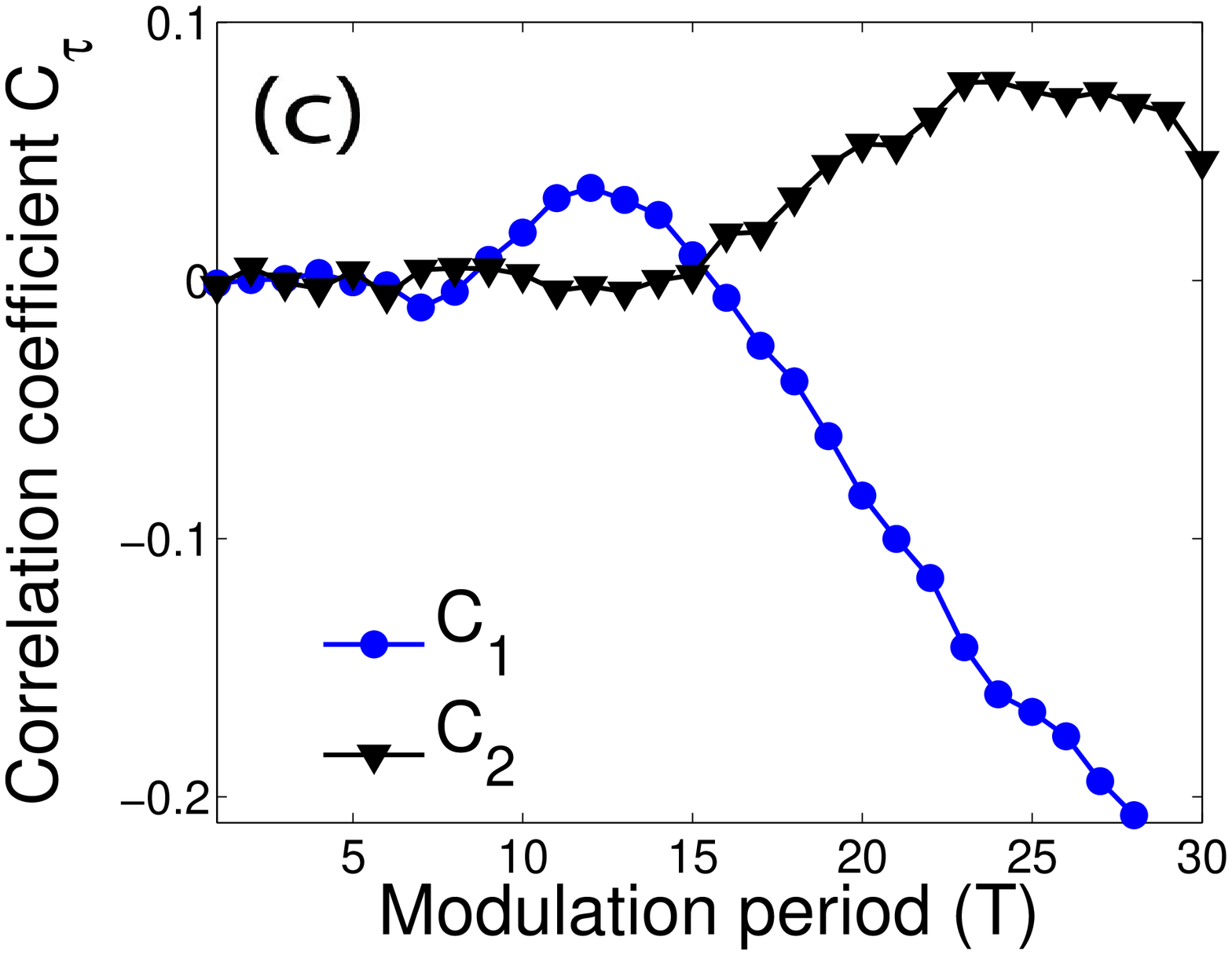}
 \includegraphics[width=2.8 cm]{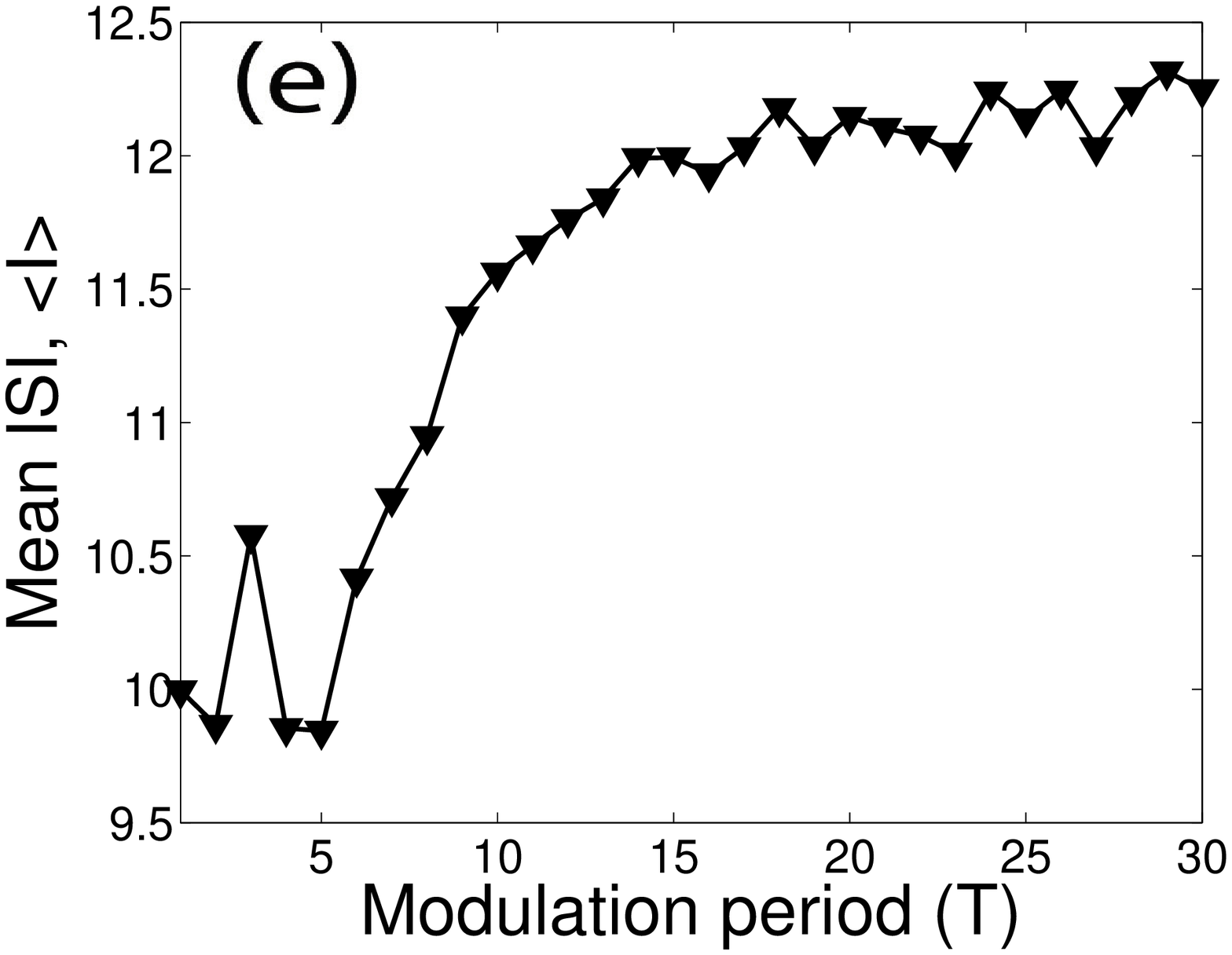}\hfill
 \includegraphics[width=2.8 cm]{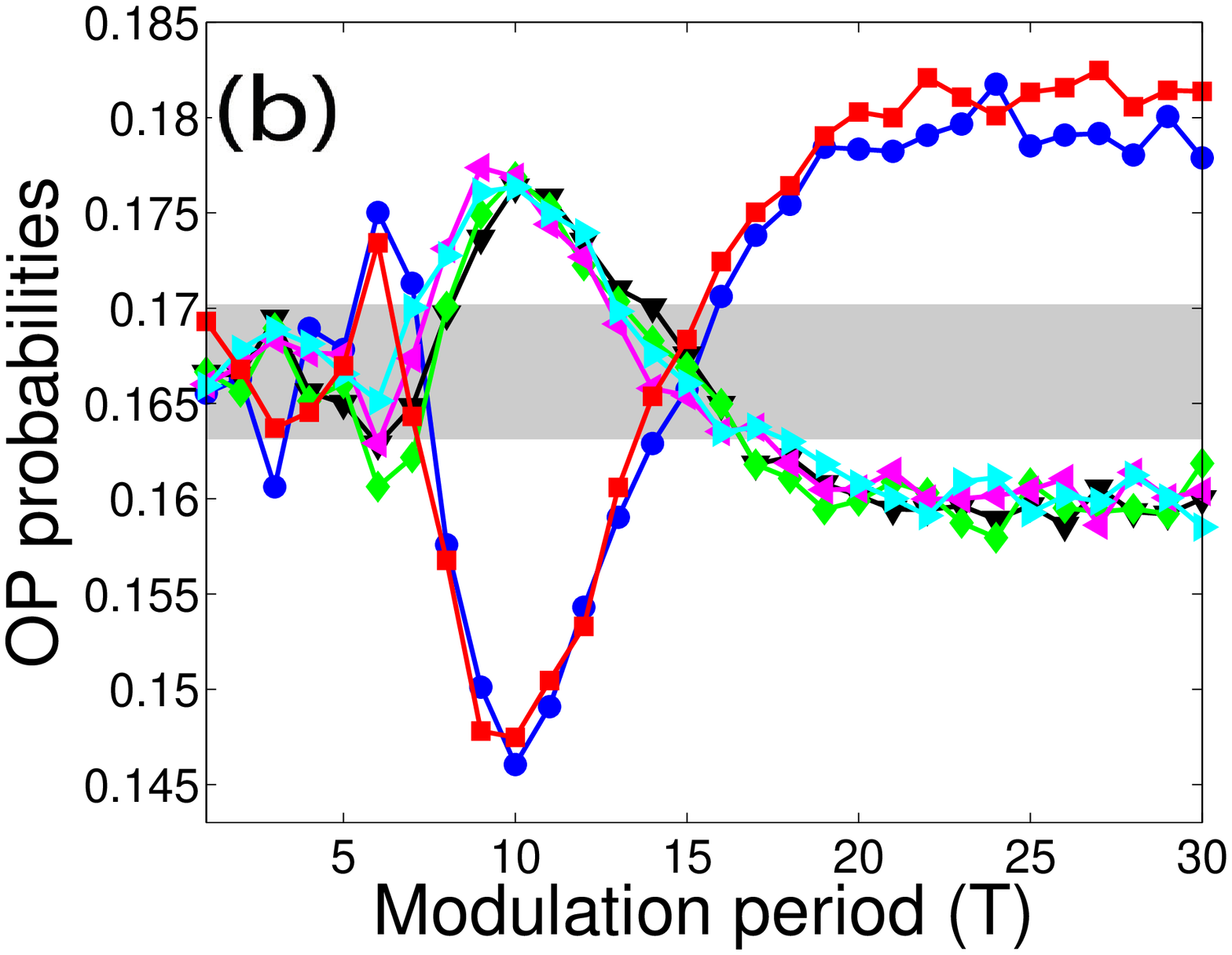}
 \includegraphics[width=2.8 cm]{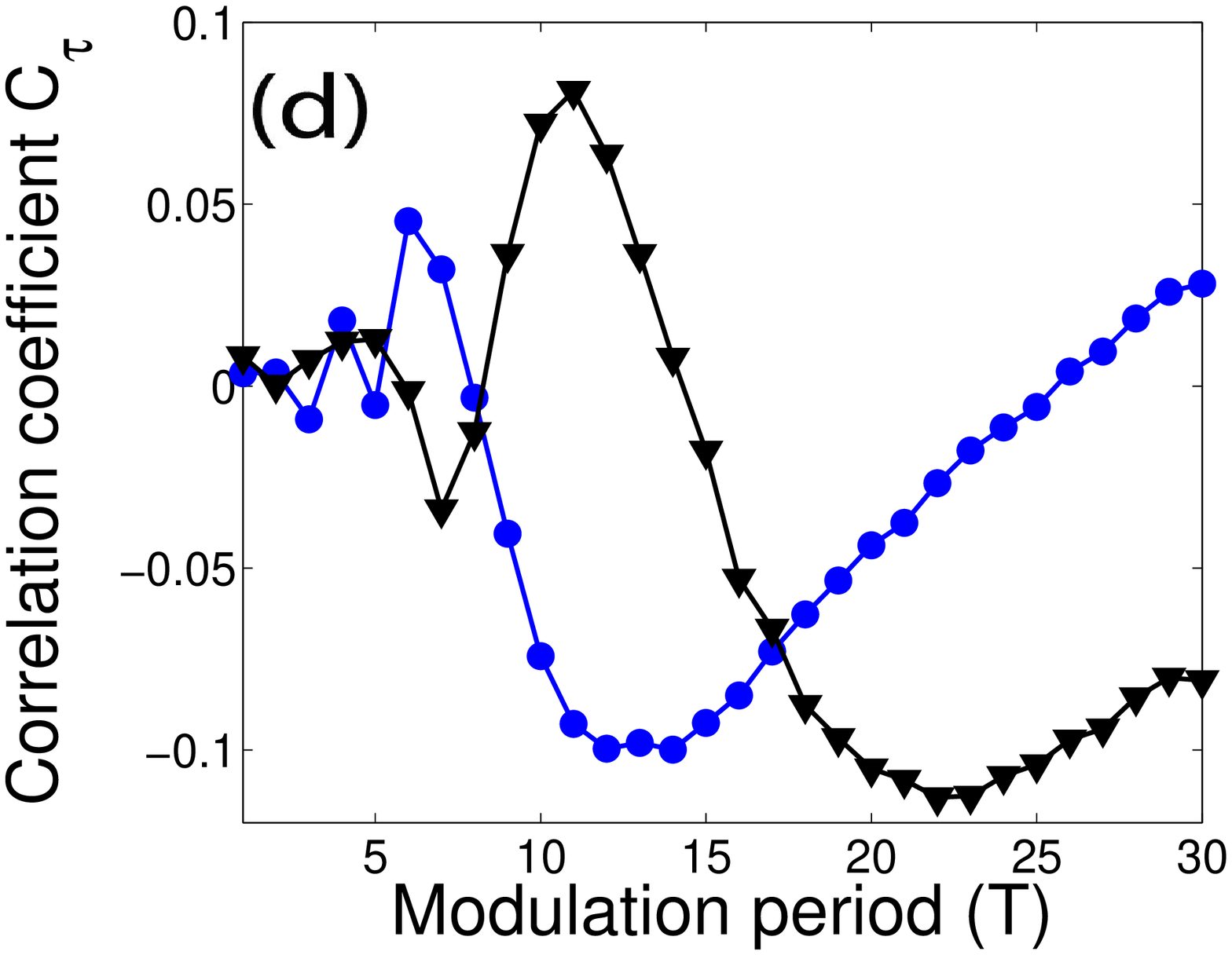}
 \includegraphics[width=2.8 cm]{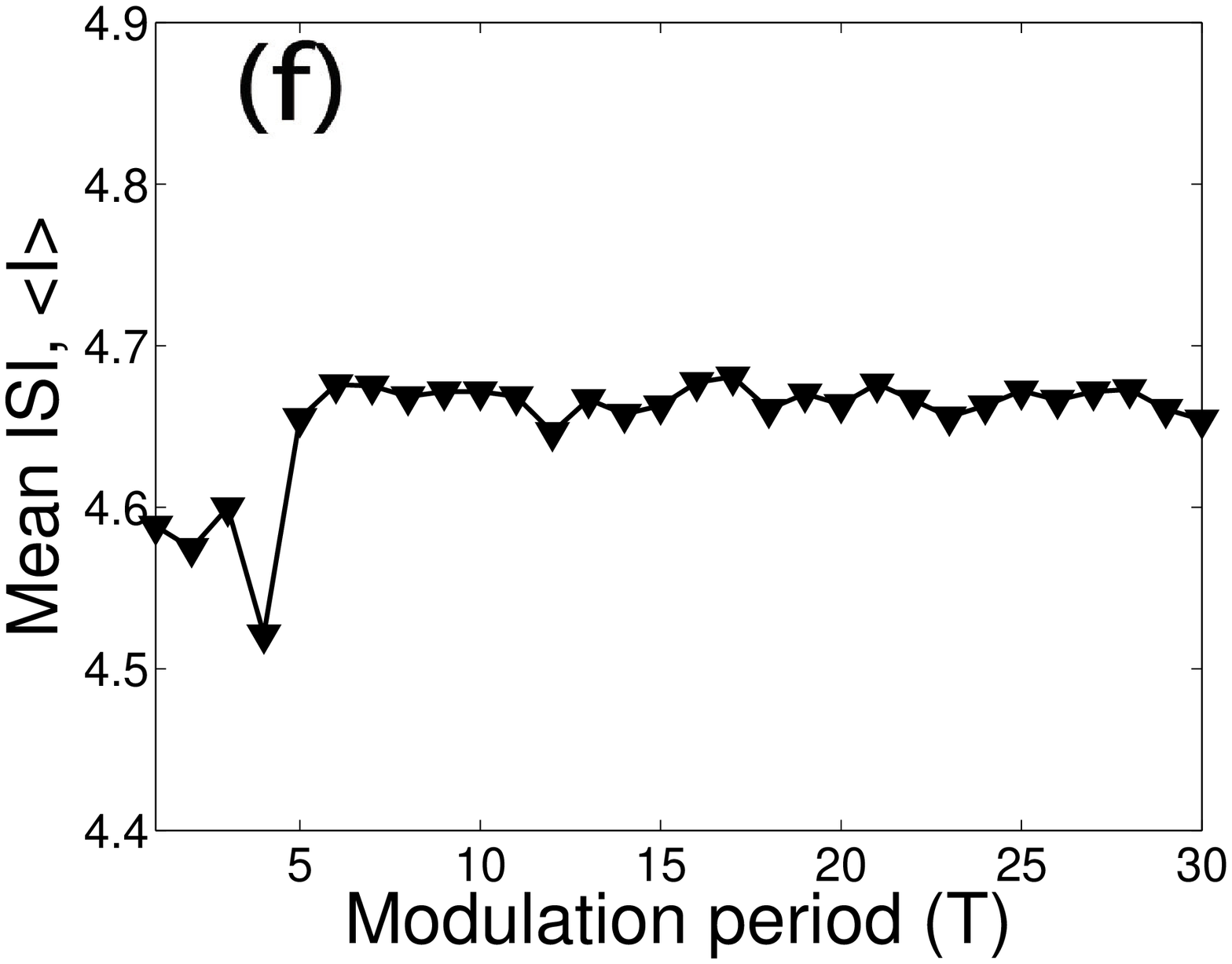}
%  
% \subfloat{\includegraphics[width=0.15\textwidth]{fig.s3a.eps}}
% \subfloat{\includegraphics[width=0.15\textwidth]{fig.s3c.eps}}
% \subfloat{\includegraphics[width=0.15\textwidth]{fig.s3e.eps}}\hfill
% \subfloat{\includegraphics[width=0.15\textwidth]{fig.s3b.eps}}
% \subfloat{\includegraphics[width=0.15\textwidth]{fig.s3d.eps}}
% \subfloat{\includegraphics[width=0.15\textwidth]{fig.s3f.eps}}
\caption{(Color online) Left column: OPs probabilities; central column: correlation coefficients and right column: mean inter-spike-interval. The  parameters are as in Fig. 4: $a_o$=0.02, (a) $D=0.015$ and (b) $D=0.035$.
\label{fig:9}}
\end{figure}
To investigate if there is a close relation between the values of the OP probabilities and the serial correlation coefficients, $C_1$ and $C_2$, and the mean ISI, $\left<I\right>$ (the inverse of the spike rate), Figs. 7-9 display, for the same parameters as Figs. 2-4, $C_1$ and $C_2$ (center column) and the mean ISI (right column). For easy comparison, the OP probabilities are also shown in the left column. 

First, we note that the variation of $\left<I\right>$ with $D$ and $T$ is not correlated to that of the OP probabilities: in particular, we see no similar trend. Second, we note that $C_1$ and $C_2$ display smooth variations, similar to those of the OP probabilities. As expected, when $C_{1}<0$ and $C_{2}>0$ the most expressed OPs are Vs and $\Lambda$s (`021', `120', `201', `102'). 

In addition, under particular conditions ``equivalent situations'' can be identified. For example, in Fig. 9 first row, for $T = 20$ and $D = 0.015$,  $\left<I\right>=12\sim T/2$. In this case, patterns `012' and `210' are the less expressed. Comparing with the second row (for $D = 0.035$), for $T=10$, $\left<I\right>=5\sim T/2$, and also patterns `012' and `210' are the less expressed.  The two situations are ``equivalent'' because in both cases $\left<I\right>\sim T/2$, and when $T = 20$ and $D = 0.015$ (Fig. 9, first row): $C_1 \sim -0.08$ and $C_2 \sim +0.05$, while when $T = 10$ and $D = 0.035$ (Fig. 9, second row), $C_1 \sim -0.08$ and $C_2 \sim +0.05$.

However, in general, no clear relations can be inferred from these plots. In order to search for such relation, in Fig. 10 we have collapsed all data sets in scatter plots, which display the OP probabilities vs. $C_1$ and $C_2$. For clarity the OP probabilities are separated in three groups: the trend patterns (`012' and `210' in the left column of Fig. 10), and the two clusters of patterns that have similar probabilities (`021' and `102' in the center column and `120' and `201' in the right column). In the scatter plots no clear relations between $C_1$ and $C_2$ and the OP probabilities are seen, but there is a well-defined trend with $C_2$ (however, the relation is not one-to-one). 

To further explore the relation between the OP probabilities and the serial correlation coefficients we have redone the scatter plots, but now selecting only significant probability values. Specifically, we consider the probability of pattern `012', $P(012)$, and instead of collapsing all data sets in Figs. 7-9, we selected only the values such that $P(012)$ is more probable or less probable than expected in the null hypothesis of equally probable OP (i.e., $P(012)$, is either above or below the gray regions in Figs. 7-9). The results are presented in Figs. 11(a) and 11(b) respectively, where the color scale indicates the value of $P(012)$. Here again we see a clear trend with $C_2$ but no trend with $C_1$.

We conclude this section by summarizing the information gained with ordinal analysis, which could not be inferred from correlation analysis:

- For a wide range of parameters, in the ISI sequences there are OPs which have almost equal probabilities: `021',`102' and `201',`120'. %}This is observed in Figs. 2(b), 3(a), and 4(a).}

- For a wide range of parameters, there is a well-defined hierarchy in the probabilities of the various OPs. For example, in Fig. 7(b), for $D > 0.04$, 

\begin{center} $P(102)=P(021)>P(120)=P(201)>P(210)>P(012)$, \end{center} 

\newpage

while in Fig. 9(a), for $T > 15$, 

\begin{center} $P(120)=P(201)>P(102)=P(021)>P(012)>P(210)$.  \end{center}
	
- The ordinal probabilities allow computing the permutation entropy, shown in Fig. 5(b), that  displays a sharp transition at $T=10$. Such transition is not seen in $\left<I\right>$, $C_1$ or $C_2$, which vary smothly with the modulation period (as shown in Fig. 9).

These observations provide a complementary approach for a qualitative comparison of empirical and synthetic ISI sequences, and can also be useful for distinguishing/classifying different types of ISI sequences.

\begin{figure*}[htbp]
\centering
% \begin{figure*}[htbp]
%  \centering
%  \includegraphics[width=16 cm]{fig.s4.eps}

 \includegraphics[width=5 cm]{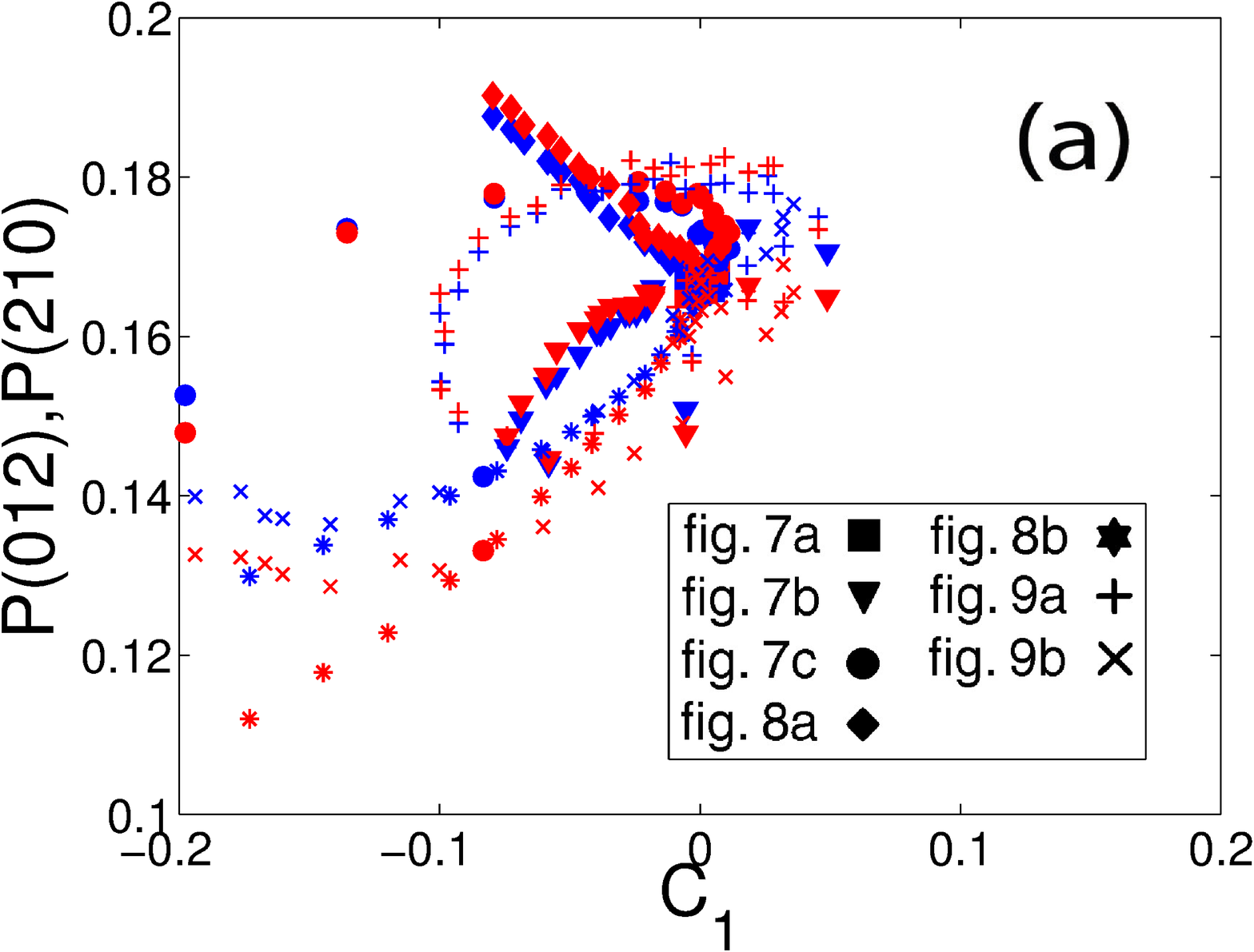}
 \includegraphics[width=5 cm]{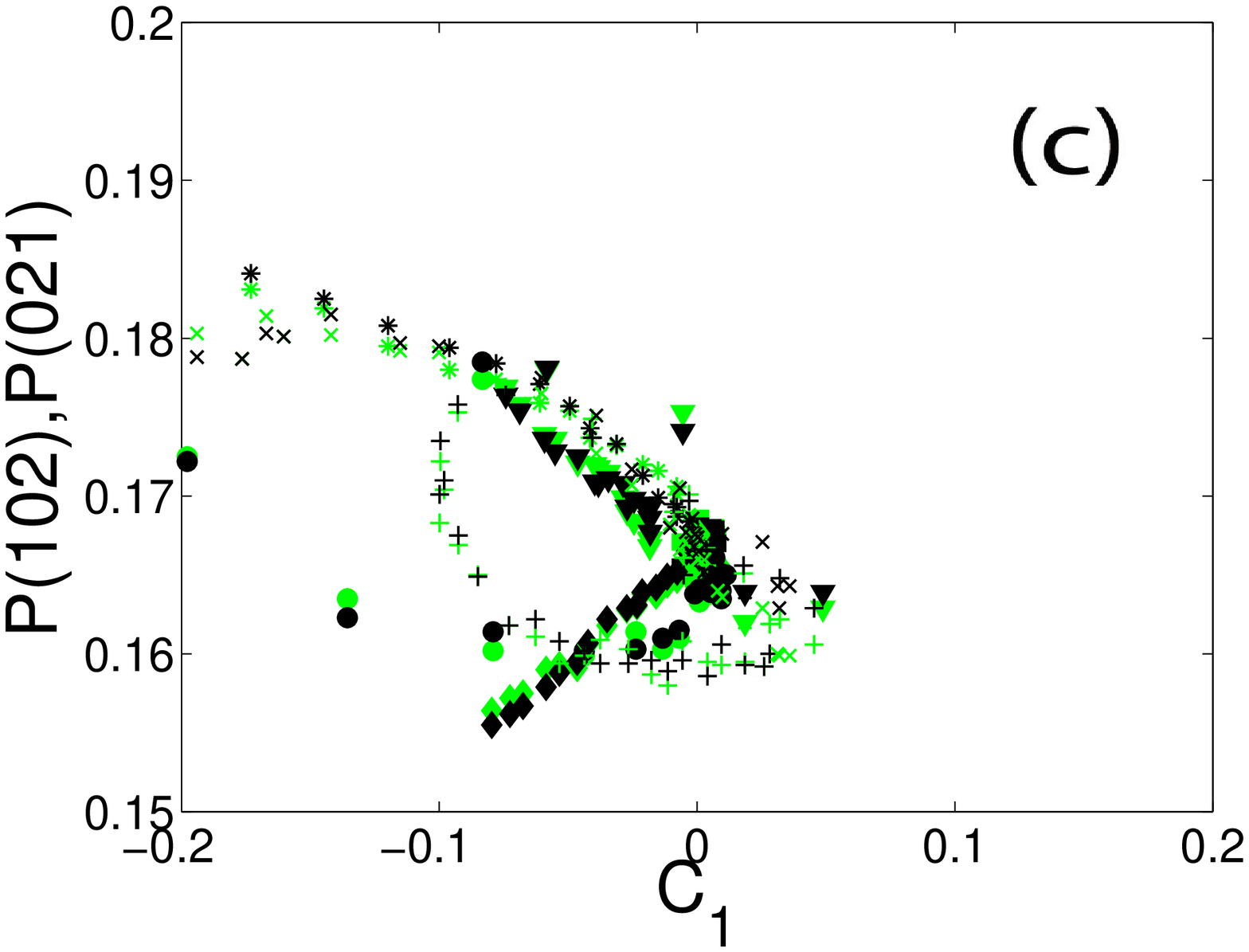}
 \includegraphics[width=5 cm]{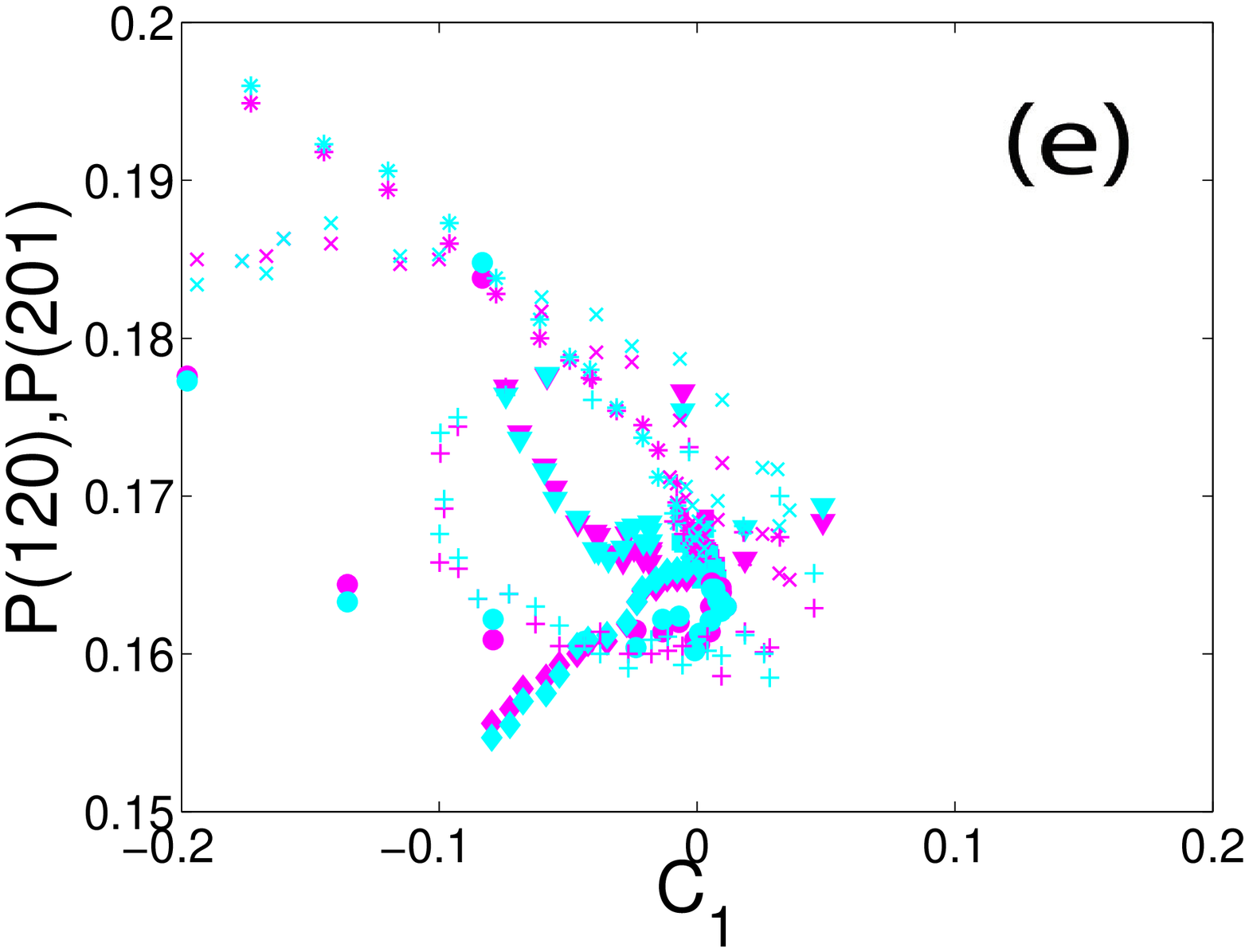}\hfill
 \includegraphics[width=5 cm]{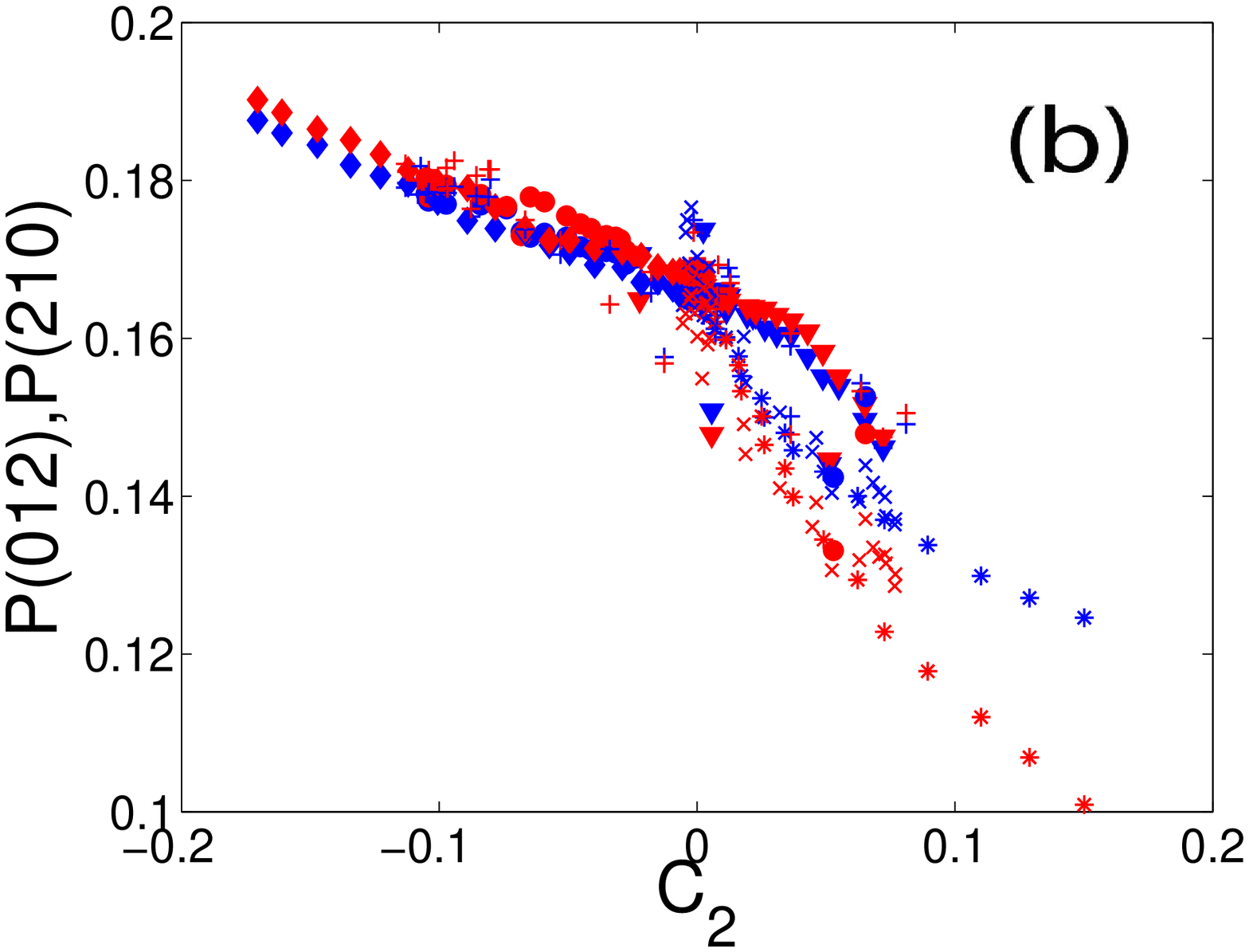}
 \includegraphics[width=5 cm]{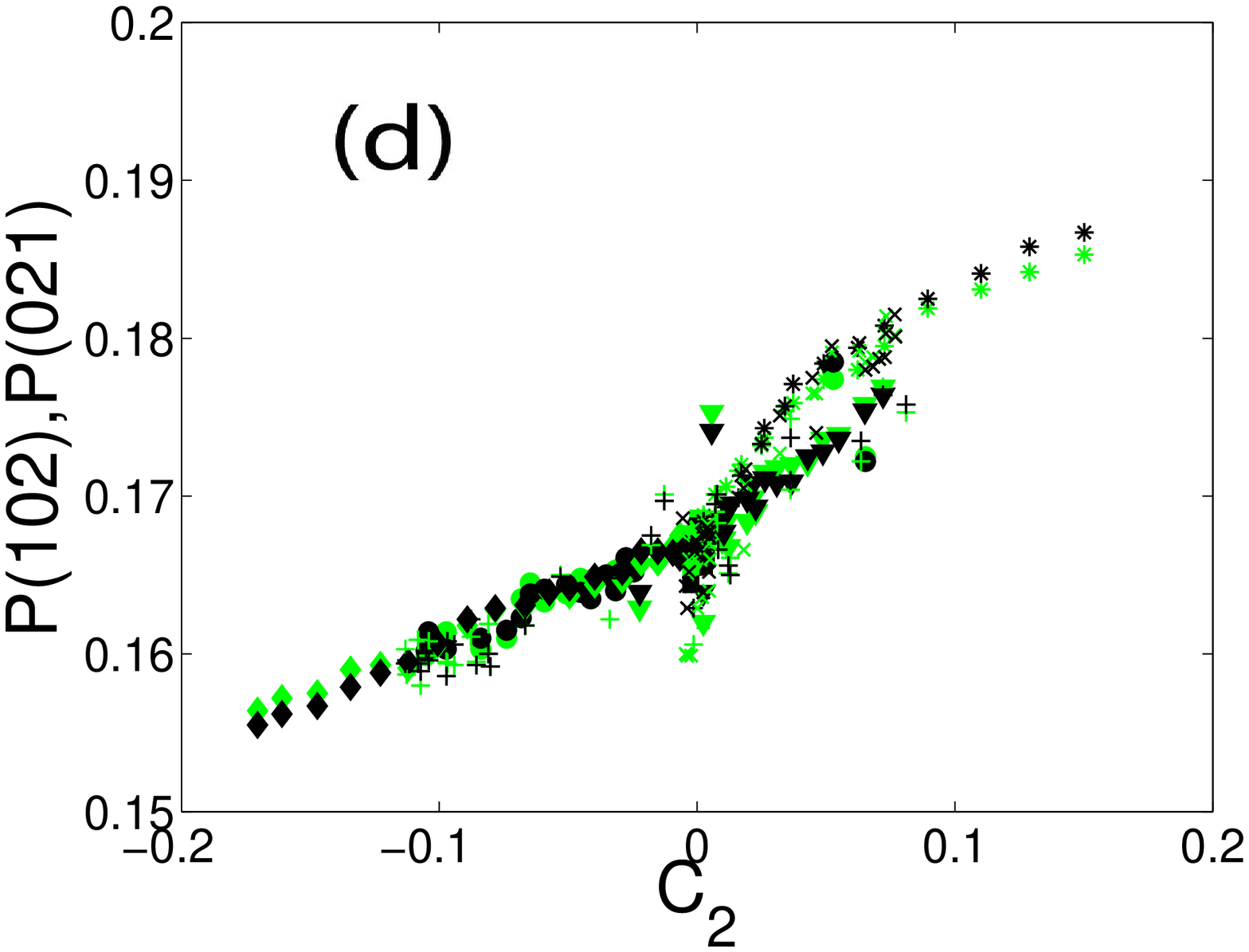}
 \includegraphics[width=5 cm]{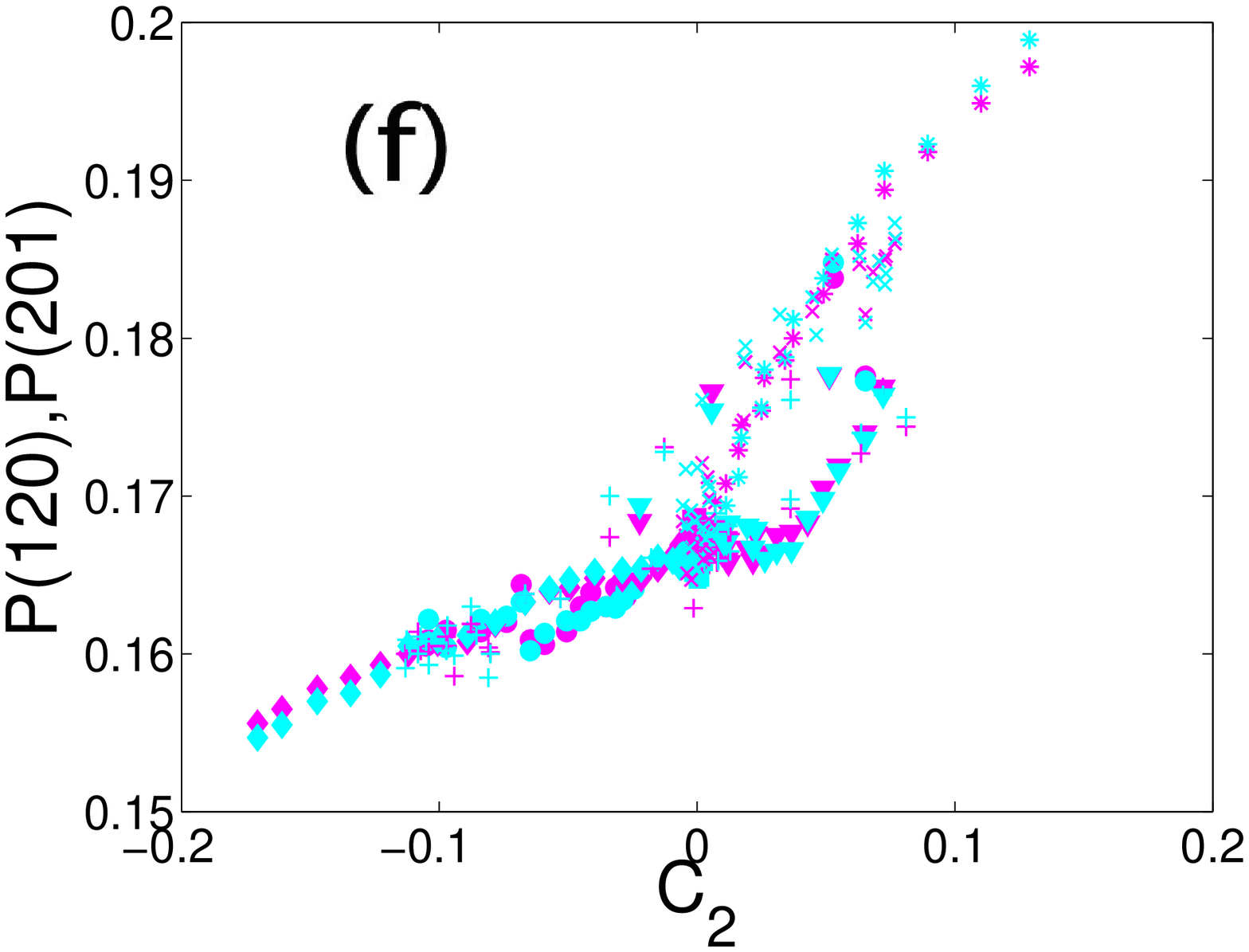}

\caption{(Color online) Scatter plots in which all the data sets shown in the three previous figures are collapsed. Here the OP probabilities are plotted vs $C_1$ (top row) and $C_2$ (bottom row). For clarity the OP probabilities are separated in three groups: the trend patterns (`012' and `210', in the left column), and the two clusters of patterns that have similar probabilities (`021' and `102' in the center column and `120' and `201' in the right column). No clear relation between $C_1$, $C_2$, and the six ordinal probabilities is seen, but there is a well-defined trend with $C_2$.
\label{fig:10}}
\end{figure*}

\begin{figure*}[htbp]
\centering
% \begin{figure*}[htbp]
%  \centering
%  \includegraphics[width=16 cm]{fig.s4.eps}
 \includegraphics[width=8 cm]{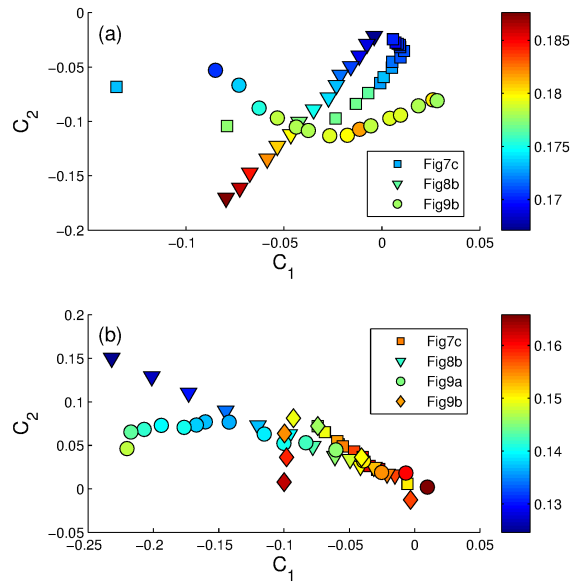}

\caption{(Color online) Scatter plots which are done by selecting only parameter values in Figs. 7-9 such that pattern `012' is more probable [panel (a)] or less probable [panel (b)] than expected in the null hypothesis of equally probable patterns (i.e., above or below the gray regions indicated in Figs. 7-9). The color code indicates the value of $P(012)$. In both panels, no clear relation between $P(012)$ and $C_1$ is seen, but there is a clear trend with $C_2$.
\label{fig:11}}
\end{figure*}

\section{Conclusions}
\label{sec4} 

To summarize, we have studied the emergence of relative temporal order in spike sequences induced by the interplay of a stochastic input and a subthreshold periodic input. By using symbolic analysis we uncovered preferred ordinal patterns, which are tuned by the period of the input signal and by the strength of the noise. We have also shown that the probabilities of specific patterns are maximum or minimum for particular values of the period of the input and the strength of the noise. Our findings could be useful for contrasting empirical and synthetic ISI sequences, for validating neuron models or estimating their parameters. Moreover, our results could motivate new experiments on single sensory neurons, to further understand the mechanisms by which they encode information about weak stimuli in noisy environments. 

This work has been supported in part by the Spanish MINECO (FIS2015-66503).


\begin{thebibliography}{99}
\bibitem{footnote} If the six OP probabilities are within the interval [$p -3\sigma$, $p + 3\sigma$], where $p=1/6$, 
$\sigma=\sqrt{{p(1-p)}/{M}}$, and $M$ is the number of OPs, then, the probabilities are consistent with the uniform distribution with 95 \% confidence level.
\bibitem{footnote2} If $I_i=I_{i+1}$, a small random value is added before computing the ordinal pattern.
\end{thebibliography}
\end{document}